\documentclass[aps,prd,amsfonts,amssymb, amsmath, showpacs, showkeys, a4paper, nofootinbib, superscriptaddress, 11pt, raggedbottom]{revtex4-2}

\usepackage{braket}
\usepackage{bm}
\usepackage{graphicx}
\usepackage{slashed}
\usepackage{subfig}
\usepackage{mathrsfs}
\usepackage{xcolor}
\usepackage{mathtools}
\usepackage[normalem]{ulem}
\usepackage{textcmds}
\usepackage{hyperref}
\usepackage{tikz}
\usepackage{float}

\def\be{\begin{equation}} 
\def\ee{\end{equation}}   
\def \eea{\end{eqnarray}}
\def \bea{\begin{eqnarray}}

\begin{document}

\title{\Large Equivalence of scalar-tensor theories and scale-dependent gravity}

\author{Philipp Neckam}
\email{philipp.neckam@tuwien.ac.at}
\affiliation{Atominstitut, Technische Universit\"at Wien,  Stadionallee 2, A-1020 Vienna, Austria}
\author{Christian K\"ading}
\email{christian.kaeding@tuwien.ac.at}
\affiliation{Atominstitut, Technische Universit\"at Wien,  Stadionallee 2, A-1020 Vienna, Austria}
\author{Benjamin Koch}
\email{benjamin.koch@tuwien.ac.at}
\affiliation{Institut f\"ur Theoretische Physik,
 Technische Universit\"at Wien,
 Wiedner Hauptstrasse 8--10,
 A-1040 Vienna, Austria}
 \affiliation{ Instituto de F\'isica, Pontificia Universidad Cat\'olica de Chile, 
Casilla 306, Santiago, Chile}
\affiliation{Atominstitut, Technische Universit\"at Wien,  Stadionallee 2, A-1020 Vienna, Austria}
\author{Cristobal Laporte}
\email{laportecristobal@gmail.com}
\affiliation{Institute for Mathematics, Astrophysics and Particle Physics (IMAPP), Radboud University,
Heyendaalseweg 135, 6525 AJ Nijmegen, The Netherlands}
\author{Mario Pitschmann}
\email{mario.pitschmann@tuwien.ac.at}
\affiliation{Atominstitut, Technische Universit\"at Wien,  Stadionallee 2, A-1020 Vienna, Austria}
\author{Ali Riahinia}
\email{ali.riahinia@tuwien.ac.at}
\affiliation{Institut f\"ur Theoretische Physik,
 Technische Universit\"at Wien,
 Wiedner Hauptstrasse 8--10,
 A-1040 Vienna, Austria}
\affiliation{Atominstitut, Technische Universit\"at Wien,  Stadionallee 2, A-1020 Vienna, Austria}
\author{Angel Rincon}
\email{angel.rincon@physics.slu.cz}
\affiliation{Research Centre for Theoretical Physics and Astrophysics, Institute of Physics, Silesian University in Opava, Bezručovo náměstí 13, CZ-74601 Opava, Czech Republic.}
\affiliation{
Instituto Universitario de Matemática Pura y Aplicada,
Universitat Polit\`ecnica de Val\`encia, Valencia 46022, Spain
}

\begin{abstract}

\vspace{12pt}
\noindent
We present a novel equivalence between scale-dependent gravity and scalar-tensor theories that have only a single scalar field with a canonical kinetic term in the Einstein frame and a conformal coupling to the metric tensor. In particular, we show that the set of well-behaved scale-dependent gravity theories can be fully embedded into scalar-tensor theories in a unique way. Conversely, there are multiple ways to write a scalar-tensor theory as a scale-dependent theory. This equivalence is established both on the level of the actions and on the level of field equations. We find that, in the context of this equivalence, the scale-setting relation $k(x)$ is naturally promoted to a dynamical field, which is made manifest by including a corresponding kinetic term in the scale-dependent action. In addition, we demonstrate that the new equivalence fits well into the framework of existing equivalences involving the aforementioned theories and $f(R)$-gravity. Finally, we apply the equivalence relations to explicit examples from both scale-dependent gravity and scalar-tensor theories.
\end{abstract}


\maketitle


\tableofcontents


\section{Introduction}

Despite the overwhelming success of general relativity (GR), evidence has accumulated over the past century that motivates searches for alternative or modified theories of gravity. In particular, questions about the nature of dark energy and dark matter remain among the most significant problems in gravitational physics and cosmology. While it is possible to account for dark energy within GR in the form of a cosmological constant, this leads to severe fine-tuning problems \cite{Weinberg:1988cp}. In addition, despite substantial theoretical and experimental efforts, the search for direct evidence of particle dark matter has so far been without success. Consequently, modified theories of gravity are extensively discussed in the literature and many different models have emerged \cite{Clifton:2011jh}. Studying the similarities, differences and interrelations of the various approaches is essential for our understanding of gravity as a whole.

In addition to the previously mentioned reasons, a key motivation for exploring theories that go beyond GR is to develop a consistent, elegant, and concise framework for describing classical and quantum phenomena in nature. However, unifying GR and quantum mechanics within a single theoretical construct remains a persistent challenge.
Thus, the limitations of GR in describing certain physical phenomena at extreme scales have, in concrete terms, motivated efforts to modify it. While GR is excellent at explaining gravity on large scales, it fails to account for quantum effects in the Planck regime, precisely where such effects become significant. This discrepancy compels us to explore modifications to GR through approaches such as quantum gravity or extended theories of gravitation. The aim is to establish a more comprehensive framework that consistently unifies gravity with the other fundamental forces while addressing outstanding empirical and theoretical challenges.

A simple way of modifying GR is by coupling an additional scalar field to the metric tensor, resulting in a so-called scalar-tensor theory (STT) of gravity \cite{Fujii:2003pa}.
This can be motivated, for example, by higher-dimensional theories whose compactified extra dimensions give rise to additional scalar degrees of freedom \cite{Wehus:2002se}. Due to the coupling of a scalar field to the metric tensor, we would expect to see an additional fifth force of Nature. However, fifth forces are already heavily constrained by Solar System-based tests \cite{Dickey1994,Adelberger2003,Kapner2007}. 
In order to circumvent such experimental bounds, some STTs are equipped with a screening mechanism that renders the fifth force feeble in environments of sufficiently large mass densities. Amongst the most prominent screened scalar field models are the chameleon \cite{Khoury2003,Khoury20032}, the symmetron \cite{Dehnen1992, Gessner1992, Damour1994, Pietroni2005, Olive2008, Brax:2010gi,Hinterbichler2010,Hinterbichler2011}, and the environment-dependent dilaton \cite{Damour1994,Gasperini:2001pc,Damour:2002nv,Damour:2002mi,Brax:2010gi,Brax:2011ja,Brax2022}. Chameleons \cite{Zaregonbadi:2025ils} and symmetrons \cite{Burrage:2016yjm,OHare:2018ayv,Burrage:2018zuj}, as well as their generalizations \cite{Kading:2023hdb}, have recently been proposed as alternatives to particle dark matter. Screened scalar field models have been tested in various experimental setups. A good overview of such experiments can be found in Ref.~\cite{Burrage:2016bwy,Brax:2021wcv}. Many tight constraints on screened scalar fifth forces stem from atom interferometry experiments \cite{Burrage:2014oza,Hamilton:2015zga,Burrage:2015lya,Elder:2016yxm,Burrage:2016rkv,Jaffe:2016fsh,Sabulsky:2018jma}, of which Ref.~\cite{Mueller2024} reports the latest results. There are also suggestions of new atom interferometry experiments \cite{Garcion:2025nnl} as well as ideas to improve existing constraints by using Bose-Einstein condensate (BEC) interferometry \cite{Hartley:2019wzu}. Other parts of the model parameter spaces have been constrained by ultra-cold neutron experiments \cite{Jenke:2014yel,Cronenberg:2018qxf,Jenke:2020obe,Fischer:2023koa,Altarawneh:2025mig}, in particular, neutron interferometry \cite{Lemmel:2015kwa,Fischer:2023eww}. As an extension, neutron Lloyd interferometry was suggested \cite{Pokotilovski:2012xuk,Pokotilovski:2013tma}. Additional constraints have been obtained from Lunar Laser Ranging \cite{Zhang:2019hos}, torsion pendulums \cite{Zhao:2022hwa}, levitated force sensors \cite{Yin:2022geb}, hydrogen-like systems \cite{Brax:2022olf}, and gravitational waves \cite{Xiong:2024vsd}. In particular, solar chameleons are actively searched for \cite{Baum:2014rka,OShea:2024jjw,Yuan:2025twx}. In this context, first measurements have been made with CAST \cite{ArguedasCuendis:2019fxj}. A recent reassessment of more general Solar System-based tests can be found in \cite{Turyshev:2025bty}. In addition, there have been multiple suggestions for other experiments that could constrain screened scalar fields. Those include mechanical systems \cite{Betz:2022djh}, quantum optomechanics \cite{Qvarfort:2021zrl,Li:2024ynr}, atomic clocks \cite{Levy:2024vyd,Levy:2024mut}, the LHC \cite{Englert:2024ryd}, terrestrial dark matter direct detection experiments \cite{Vagnozzi:2021quy}, and Casimir force experiments \cite{Klimchitskaya:2024dvk,Rene}. For the last (and related experiments), there exist analytical studies with exact solutions for the fields \cite{Ivanov:2016rfs,Brax:2017hna,Pitschmann:2020ejb}. Extensions of Casimir experiments by taking into account the dynamical Casimir effect \cite{Baez-Camargo:2024jia} or quantum and thermal pressures \cite{Brax:2018grq,Fischer:2024gni} have also been discussed. Accounting for open quantum dynamical effects might, in the future, lead to even tighter bounds on screened scalar fields \cite{Burrage:2018pyg,Kading:2023mdk,Kading:2024jqe}. In order to optimize experiments searching for screened scalar fields, there are numerical methods \cite{Fischer:2024coj} and machine learning approaches \cite{Briddon:2023ayq}. In particular, there is SELCIE \cite{Briddon:2021etm,Briddon:2023ipy}, a tool that allows for studying field profiles around arbitrary sources. In astrophysics and cosmology, there have been suggestions to obtain constraints on screened scalar fields from galaxy clusters \cite{Pizzuti:2024hym,Butt:2025pdx}, gravitational redshift measurements \cite{Hu:2025uqb} or from non-local quantum correlations between entangled spin pairs \cite{Feleppa:2025clx}. There have a also been studies of screened scalar fields around white dwarfs \cite{Bachs-Esteban:2025mxl,Bachs-Esteban:2024vym}, in NFW halos \cite{Tamosiunas:2021kth} and in cosmic voids \cite{Tamosiunas:2022tic}. For symmetrons, in particular, the fullshape galaxy power spectrum \cite{Morales-Navarrete:2025lzp} and a connection to cosmic strings \cite{Nezhadsafavi:2025pzg} have been discussed. It has also been suggested to study spacetimes with conformally coupled screened scalar fields in BEC analogue gravity simulations \cite{Hartley:2018lvm}, and Ref.~\cite{SevillanoMunoz:2024ayh} provides a particle physics perspective on screening.
The most up-to-date constraints on all three models are provided by Refs.~\cite{Fischer:2024eic,Brax:2018iyo,Burrage:2017qrf}. However, note that recently it has been suggested that taking into account quantum corrections to symmetrons might significantly weaken their induced fifth forces \cite{Udemba:2025csd}.

Another modified gravity approach is scale-dependent (SD) gravity which is inspired by asymptotic safety \cite{Weinberg:1980gg,Reuter:2019byg}. 
The scale-dependent formalism in gravity has its roots in several seminal works. A pioneering contribution was made by Reuter and Weyer, who explored the concept of renormalization group-improved gravitational actions \cite{Reuter:2003ca}. A detailed exposition of the underlying framework, namely the Functional Renormalization Group, can be found in the textbook {\it Quantum Gravity and the Functional Renormalization Group} by Reuter and Percacci \cite{Reuter:2019byg}. The potential implications of quantum Einstein gravity on astrophysical scales are discussed in \cite{Reuter:2004nx}, while the effects of running couplings on galactic rotation curves are analyzed in \cite{Reuter:2004nv}.
A concise introduction to these ideas is provided by Percacci \cite{Percacci:2017fkn}.
These developments established the basis for what is now known as scale-dependent gravity. The first explicit application of this framework appeared in \cite{Koch:2010nn}, with subsequent progress at the cosmological level. Further exploration of the phenomenological consequences of this approach can be found in \cite{Koch:2014joa} and \cite{Contreras:2016mdt}.
In this approach, one works with a low-energy effective action that is obtained by integrating out all quantum fluctuations above a certain scale $k(x)$ in the full (unknown) action. The ignorance of the precise UV-structure of the theory then manifests as a running of the gravitational couplings. While the effective action generally consists of infinitely many terms, one usually works with the so-called Einstein-Hilbert truncation, which amounts to standard GR with the gravitational couplings being replaced by functions of $k$. A major advantage of this approach is that it allows to assess quantum properties of gravity without having a concrete UV-completion at hand. Note that there have also been other recent discussions on the physical running of the Newton coupling and the cosmological constant \cite{Kawai:2025wkp} with applications to black hole thermodynamics \cite{Chen:2022xjk} and cosmological models \cite{Chen:2024ebb,Chen:2025ybu}.

Finally, there is also the theory of $f(R)$-gravity \cite{Sotiriou:2008rp}, in which the Ricci scalar in the Einstein-Hilbert action is replaced by an arbitrary function $f(R)$ thereof. For $f(R)$-gravity, it is well-known that it can be embedded into both SD gravity \cite{Calzada:2023yiq} and STTs \cite{Nojiri:2017ncd}. Note that the latter equivalence also holds at the quantum level on-shell, but is broken by off-shell quantum corrections \cite{Ruf:2017xon,Ohta:2017trn}. However, only specific SD theories and STTs can be expressed as $f(R)$-gravity theories. These relations suggest that there exists also a third equivalence, one between SD gravity and STTs themselves. While special cases of such an equivalence have been encountered in the literature, e.g. in Ref.~\cite{Reuter:2003ca}, the general case has not been investigated yet. In this work, we will bridge this gap by providing a detailed account of the full equivalence b-etween these two theories. This is done both on the level of the actions and on the level of the field equations. The resulting equivalence relations are simple and can readily be applied to any model satisfying some well-motivated consistency conditions.

The outline of this article is as follows. We start by briefly reviewing STTs and SD gravity in Sec.~\ref{sec:STTSD}. In Sec.~\ref{sec:EquiAct}, we give a detailed analysis of the equivalence between the two theories on the level of the actions. More precisely, we determine all possible ways in which one of the theories can be reformulated as a theory of the other type and give precise relations that allow us to obtain the equivalent theory from the original one. Sec.~\ref{sec:Consistency} is dedicated to checking the consistency of the derived equivalence relations by mapping $f(R)$-gravity onto itself. In Sec.~\ref{sec:FieldEqs}, we demonstrate the validity of the equivalence relations obtained in Sec.~\ref{sec:EquiAct} on the level of the field equations. In particular, we show that the field equations of one theory are mapped onto those of the equivalent theory by the derived relations. This also reveals an important relation between the equations of motion for the scale-setting relation $k$ and the scalar-field $\phi$. Sec.~\ref{sec:Einstein} briefly describes the equivalence in the Einstein-frame, which is very useful for practical applications. Subsequently, in Sec.~\ref{sec:Examples}, we apply the equivalence relations to concrete models of both STT and SD gravity. Finally, we end the article with a discussion of the results and an outlook on future research directions.


\section{Scalar-tensor theories and scale-dependent gravity}
\label{sec:STTSD}

In this section, we provide brief introductions to the mathematics of STTs and SD gravity necessary for the present article. 


\subsection{Scalar-tensor theories}

STTs of gravity are theories in which additional scalar degrees of freedom are coupled to the metric tensor. While there can be multiple scalar fields in an STT and, generally, the kinetic structure of such models can be rather intricate 
(see Horndeski \cite{Horndeski:1974wa} and beyond Horndeski theories \cite{Gleyzes:2014dya}), 
we restrict our discussion to cases with only a single conformally coupled scalar field $\phi$ that has a canonical kinetic term in the so-called Einstein frame, which is one of the two conformal frames prevalently used to formulate STTs \cite{Fujii:2003pa}. Note that the equivalence of conformal frames is still an open question; see, for example, Ref.~\cite{Azri:2018gsz}.
In principle, we could also consider STTs with more complicated kinetic structures, but we expect that such theories would lead to more intricate equivalence relations, whose discussion is beyond the scope of the present article. In the Einstein frame, a STT can be described by the action
\begin{eqnarray} 
\label{SI}
S_\text{STT} &=& \int d^4x \sqrt{-g} \left\{ \frac{M_P^2}{2}R - \frac{1}{2}(\partial \phi)^2 - V(\phi) \right\}~,
\end{eqnarray}
where $M_P$ is the Planck mass, $R$ is the Ricci scalar and $V(\phi)$ describes self-interactions of the scalar field $\phi$. Note that, unless indicated otherwise, we do not consider the coupling of matter to the theory in this work.

Using a conformal transformation
\begin{eqnarray}  
\label{ctrafo}
g_{\mu \nu} &=& A^{-2}(\phi) \Tilde{g}_{\mu \nu} ~,
\end{eqnarray}
with the conformal factor $A^2(\phi)$, we can go from the Einstein to the Jordan frame whose quantities are labeled by $\Tilde{~}$. The inverse metric and the metric determinant transform as $g^{\mu \nu} = A^2 \Tilde{g}^{\mu \nu}$ and $\sqrt{-g} = A^{-4} \sqrt{-\Tilde{g}}$, respectively, while for the Ricci scalar we have
\begin{eqnarray}
\label{cRicci}
R &=& A^2 \left(\Tilde{R}+6 \Tilde{\Box} \ln A - 6(\Tilde{\partial} \ln A)^2 \right)~,
\end{eqnarray}
where $(\Tilde{\partial} \ln A)^2 := \Tilde{\partial}_\alpha \ln A \ \Tilde{\partial}^\alpha \ln A $. Note that for an arbitrary function we have $(\Tilde{\partial} f)^2 = A^{-2} (\partial f)^2$. Consequently, for the Jordan frame action we find
\begin{eqnarray}
\Tilde{S}_\text{STT} &=& \int d^4x \sqrt{-\Tilde{g}} \left\{ \frac{M_P^2}{2} A^{-2} \left[ \Tilde{R}+6 \Tilde{\Box} \ln A - 6(\Tilde{\partial} \ln A)^2 \right] -\frac{1}{2}A^{-2}(\Tilde{\partial} \phi)^2 -A^{-4}V(\phi) \right\}~,
\end{eqnarray}
which can be rewritten as \cite{MarioHabil}
\begin{eqnarray}
\label{SIJ}
\Tilde{S}_\text{STT} &=& \int d^4x \sqrt{-\Tilde{g}} A^{-2} \left\{ \frac{M_P^2}{2} \Tilde{R} -\frac{1}{2} \kappa^2(\phi)(\Tilde{\partial} \phi)^2-A^{-2}V(\phi) \right\} ~,
\end{eqnarray}
with the modified kinetic term given by
\begin{eqnarray} 
\label{kappa}
\kappa^2(\phi) &:=& 1 - \frac{3}{2} M_P^2 \left(\frac{\partial \ln A^{-2}}{\partial \phi} \right)^2~.
\end{eqnarray}


\subsection{Scale-dependent gravity}

Here, we will briefly explain the fundamental ingredients and principles of SD gravity. There are many attempts to incorporate quantum effects in the gravitational sector. More precisely, within quantum-inspired gravitational theories, quantum corrections have been consistently incorporated into classical backgrounds, leading to fruitful developments like SD gravity. In general, there are three main approaches to introducing quantum corrections into classical gravitational solutions \cite{Reuter:2003ca,Rincon:2022hpy}: at the level of the solutions, at the level of the equations of motion, or at the level of the actions.
One of the known and well-understood quantum effects that arises in quantum field theories is the running behavior of the fundamental couplings of the theory.
While deriving SD theories from renormalization–group approaches can be questioned~\cite{Donoghue:2019clr,Donoghue:2024uay}, the appearance of effective scale-dependence is a ubiquitous phenomenon. It is expected to arise in a wide range of underlying theories, including the low–energy limit of string theory~\cite{Hur:2025lqc}.

In the framework of SD gravity, we promote the fundamental couplings of the theory to running couplings which evolve with some scale $k$, i.e., $\{a,b,c,...\} \to \{a(k),b(k),c(k),...\}$. In addition, we can also realize other terms with the same scale-dependence as the fundamental couplings in the effective average action of SD gravity such as SD kinetic terms. Note that, in general, $k$ can depend on spacetime coordinates.
Introducing quantum effects at the level of the action, such that the equations of motion derived from the action already contain these quantum features, the action of SD gravity used in this article is given by
\begin{eqnarray} 
\label{SIIIJ}
\Tilde{S}_\text{SD} 
&=& 
\int d^4x \sqrt{-\Tilde{g}} \left( \frac{\Tilde{R} - 2\Lambda(k)}{2G(k)} -B(k)(\Tilde{\partial} k)^2 \right)~,
\end{eqnarray}
where the Newton constant $G$ and the cosmological constant $\Lambda$ have been replaced by functions depending on the scale-setting relation $k(x)$. The SD equations of motion are obtained by varying the action with respect to the metric. Since the Ricci scalar is coupled to the scale-dependent Newton constant via a term $\frac{\Tilde{R}}{G(k)}$, we interpret Eq.~(\ref{SIIIJ}) as the Jordan frame action of the theory. Note that in practically all SD gravity models $B(k)=0$ is used \cite{Rincon:2019ptp,Alvarez:2022wef,Alvarez:2022mlf,Alvarez:2023ywi,Rincon:2022hpy}. Furthermore, the reader should note that the absence of the kinetic term in the action plays a fundamental role in the formalism of SD gravity, as it enables new families of solutions depending on the functional form of Newton’s coupling. It is also important to emphasize that this absence is critical because, conventionally, it implies that the auxiliary scalar field 
$k(x)$ is non-dynamical. In the present study, however, by including the kinetic term, we explore a significantly broader sector of the theory and its corresponding solution space. In particular, in the context of an equivalence to STTs, it makes sense to consider a kinetic term for the scale $k(x)$ with an unspecified $B(k)$.
Currently, there is no unique way of determining the scale-setting relation $
k(x)$. However, progress has been made in finding physically motivated conditions to fix the scale \cite{Koch:2014joa,Alvarez:2023ywi,Alvarez:2022wef}. We will return to this issue in Sec.~\ref{sec:FieldEqs}.


\section{Equivalences}
\label{sec:Equivalences}

In this section, we will discuss the equivalence between SD gravity and STTs on different levels. At first, we will establish it at the level of the actions in Sec.~\ref{sec:EquiAct}. In Sec.~\ref{sec:Consistency}, we will then check the consistency of the found equivalence relations by connecting them to known results from the literature, such that we can use them to map $f(R)$-gravity onto itself. Investigating the equivalence on the level of field equations in Sec.~\ref{sec:FieldEqs} will unravel an interesting connection between the equations of motions in SD gravity and STTs. Finally, for practical purposes, we will also discuss the equivalence in the Einstein frame; see Sec.~\ref{sec:Einstein}.


\subsection{Equivalence on the level of the actions}
\label{sec:EquiAct}

At first, we study the equivalence of STTs and SD gravity on the level of the actions. For our discussion, we assume that $\Lambda(k)$ does only depend on $k$ and not on any derivatives of the scale-setting parameter, which is reasonable since the prefactor of the kinetic term in the SD theory is unspecified, which renders allowing for a kinetic term in $\Lambda$ unnecessary. Similarly, we make the assumption that $V(\phi)$ is only dependent on $\phi$, not on any derivatives. This ensures that the form of the kinetic term in a STT cannot be arbitrarily changed.

Since the SD action in Eq.~(\ref{SIIIJ}) is in the Jordan frame, we want to compare it to the STT Jordan frame action in Eq.~(\ref{SIJ}) in order to find conditions under which the actions coincide, i.e., we demand
\begin{eqnarray}
\label{Jequiv}
\frac{\Tilde{R}}{2G(k)} -B(k)(\Tilde{\partial} k)^2 -\frac{\Lambda(k)}{G(k)}  &\stackrel{!}{=}&  \frac{M_P^2}{2A^2(\phi)} \Tilde{R} -\frac{1}{2A^2(\phi)} \kappa^2(\phi)(\Tilde{\partial} \phi)^2-\frac{V(\phi)}{A^4(\phi)}~.
\end{eqnarray}
If this equivalence happens to be the case, 
there must exist a field redefinition such that we  either have $k(\phi(x))$ or $\phi(k(x))$ in order to translate one side of this expression into the other. Note that this does not mean that the SD theory had any a priori dependence on the field $\phi$, it only means that after the scale-setting, the effect of SD became tree-level indistinguishable from the effect of this particular classical field $\phi$.
As a consequence, together with the assumptions made above, this leads us to the following relations:
\begin{eqnarray} \label{Aequiv}
G &=& \frac{A^2}{M_P^2}~,
\\
 \label{Bequiv}
B(k)(\Tilde{\partial} k)^2 &=& \frac{1}{2A^2} \kappa^2(\phi)(\Tilde{\partial} \phi)^2~,
\\
\label{Vequiv}
\frac{\Lambda}{G} &=& \frac{V}{A^4}~.
\end{eqnarray}
In what follows, we will analyze the two possible directions separately.


\subsubsection{Mapping SD gravity to STTs}
\label{sssec:SDtoSTT}

When starting from a SD gravity theory, we assume that a scale-setting relation $k(x)$ has been fixed and that the functions $G(k(x))$, $B(k(x))$ and $\Lambda(k(x))$ are known. These quantities can then be used to define $\phi$, $A^2(\phi)$ and $V(\phi)$ using the relations (\ref{Aequiv}) - (\ref{Vequiv}), which then determine the STT. 
Since $G$, $B$ and $\Lambda$ can only depend on $x$ via the scale-setting relation $k(x)$, all expressions in the STT corresponding to these functions must satisfy this requirement as well. From this, we conclude that $\phi(x)\equiv \phi(k(x))$. This is similar to the equivalence of SD gravity and $f(R)$-theory \cite{Calzada:2023yiq}, where it is necessary to ensure that the function $f$ defined from the SD quantities is indeed only a function of the Ricci scalar. 

From Eqs.~(\ref{Aequiv}) and (\ref{Vequiv}) we easily find
\begin{eqnarray}
\label{A2aA1}
A^2(\phi(k)) &=& M_P^2 G(k) ~,
\end{eqnarray}
and
\begin{eqnarray}\label{A2aV1}
V(\phi(k)) &=& M_P^4 G(k)\Lambda(k)~.
\end{eqnarray}
The kinetic term of the STT becomes
\begin{eqnarray}
\kappa^2(\Tilde{\partial}\phi)^2 &\rightarrow& \kappa^2\left( \frac{\partial \phi}{\partial k} \right)^2(\Tilde{\partial}k)^2~,
\end{eqnarray}
such that Eq.~(\ref{Bequiv}) gives
\begin{eqnarray}
\label{kin1}
B(k) = \frac{1}{2M_P^2G(k)}\kappa^2\left( \frac{\partial \phi}{\partial k} \right)^2~.
\end{eqnarray}
In the special case $B(k)=0$, we find $\kappa^2 =0$ and
\begin{eqnarray}
\label{eq:A2SDasSTT}
A^2(\phi(k)) &=& \exp^{\pm\sqrt{\frac{2}{3}}\frac{1}{M_P}\phi(k(x))}~,
\end{eqnarray}
which is the same scale factor obtained when translating $f(R)$-gravity into a STT \cite{Sotiriou:2008rp,Kading:2019vyb}. This does not come as a surprise since we already know that there is an equivalence between $f(R)$-gravity and a SD gravity with a vanishing kinetic term \cite{Calzada:2023yiq}. This indicates that the equivalence found here complements the existing equivalence relations. We will elaborate more on this in Sec.~\ref{sec:Consistency}. For a general $B(k)$, we rewrite Eq.~(\ref{kin1}) as 
\begin{eqnarray}
\label{ODE}
\left( \frac{\partial \phi}{\partial k} \right)^2 
&=&
2M_P^2 G(k) B(k) +\frac{3}{2}\frac{M_P^2}{G(k)^2}\left( \frac{\partial G}{\partial k} \right)^2~,
\end{eqnarray}
which is solved by
\begin{eqnarray}
\label{A2a3}
\phi(k)
&=&
\pm \int dk \left[2M_P^2G(k)B(k) + \frac{3}{2}\frac{M_P^2}{G(k)^2}\left( \frac{\partial G}{\partial k} \right)^2 \right]^{\frac{1}{2}}~.
\end{eqnarray}

We learn several things from Eqs.~(\ref{A2aA1}), (\ref{A2aV1}) and (\ref{A2a3}). First, we see that, for a given SD gravity theory, these relations define an, up to a sign, unique STT. Second, by investigating for which functions $G$ and $B$ Eq.~(\ref{A2a3}) yields an analytic solution for $\phi$, we can give precise conditions for a SD gravity theory to be equivalent to a STT. More precisely, we need $G(k)$ and $B(k)$ to be smooth in order to arrive at a smooth $\phi(k)$. In addition, we must require $\forall x\!\!:G(k(x)) \neq 0$ to avoid singularities. Together with the smoothness requirement for $G$ and the fact that the Newton constant is positive, this results in $G(k) > 0$. Finally, since we need the argument of the square root to be non-negative and want $\phi$ to be invertible on the image of $k$, we require 
\begin{eqnarray}
\label{eq:third}
B(k) &>& -\frac{3}{4} \frac{1}{G(k)^3}\left( \frac{\partial G(k)}{\partial k} \right)^2~.    
\end{eqnarray}
Together with the monotonicity of the square root, this ensures global invertibility of $\phi(k)$. The first requirement is automatically satisfied for any healthy SD theory, while the second requirement implies that gravity must remain attractive at any scale. Though, the most interesting one is the last requirement since it shows that the prefactor of the kinetic term in SD gravity can actually be negative. Note that the right-hand side of Eq.~(\ref{eq:third}) is always negative, which means that theories with $B(k)=0$, as well as theories with a kinetic term that has the correct sign, automatically satisfy this requirement.

Since any well-behaved SD gravity model discussed in the literature satisfies the requirements stated above, see Ref.~\cite{Rincon:2019ptp} and references therein, we have shown that this class of theories can be fully embedded into STTs. Next, we will see that the situation is different in the opposite direction.


\subsubsection{Mapping STTs to SD gravity}
\label{sec:STT2SD}

When starting from a known STT and working toward an equivalent SD gravity, we assume that the functions $\phi(x)$, $A^2(\phi(x))$ and $V(\phi(x))$ are given and can be used to define $k$, $G(k)$, $\Lambda(k)$ and $B(k)$. Since $A$ and $V$ can only depend on $x$ via the scalar field $\phi(x)$, all expressions in the SD theory corresponding to these functions must satisfy this requirement as well, resulting in $k(x)\equiv k(\phi(x))$.

From Eqs.~(\ref{Aequiv}) and (\ref{Vequiv}) we obtain
\begin{eqnarray}
\label{B3G}
G(k(\phi)) &=& \frac{1}{M_P^2}A^2(\phi) ~,
\end{eqnarray} 
and
\begin{eqnarray}
\label{B3Lambda}
\Lambda(k(\phi)) &=& \frac{V(\phi)}{M_P^2A^2(\phi)}~.
\end{eqnarray} 
In order to express the right-hand sides of these equations in terms of $k$, we need to require that $k(\phi)$ is invertible on the image of $\phi(x)$, which is fine since $k$ takes a scalar function as an argument.
The kinetic term of the SD theory becomes
\begin{eqnarray}
B(k) (\Tilde{\partial} k)^2 &\rightarrow& B(k(\phi))\left( \frac{\partial k(\phi)}{\partial \phi} \right)^2 (\Tilde{\partial}\phi)^2~,
\end{eqnarray} 
i.e., we obtain a kinetic term for $\phi$ that can match that of the STT. More precisely, we need to have
\begin{eqnarray}
\label{kinetic2}
B(k(\phi))\left( \frac{\partial k(\phi)}{\partial \phi} \right)^2 &=& \frac{1}{2M_P^2 G(k(\phi))}-\frac{3}{4 G(k(\phi))^3} \left( \frac{\partial G(k(\phi))}{\partial k(\phi)} \right)^2 \left( \frac{\partial k(\phi)}{\partial \phi} \right)^2~.
\end{eqnarray} 
If we do not make any assumptions about the specific form of $k(\phi)$, we see that $B(k(\phi))$ must actually contain a contribution that cancels the term $\left( \frac{\partial k(\phi)}{\partial \phi} \right)^2$ in order to reproduce the first term on the right-hand side of Eq.~(\ref{kinetic2}). Thus, not only do we need to demand that $k(\phi)$ is invertible on the image of $\phi(x)$, the derivative of $k$ must satisfy the same requirement. For every $k(\phi)$ satisfying these requirements, we can then introduce a corresponding $B(k(\phi))$
\begin{eqnarray}
\label{Bsolution}
B(k(\phi)) &=& \left( \frac{\partial k(\phi)}{\partial \phi} \right)^{-2} \frac{1}{2M_P^2 G(k(\phi))}-\frac{3}{4 G(k(\phi))^3} \left( \frac{\partial G(k(\phi))}{\partial k(\phi)} \right)^2~,
\end{eqnarray} 
which defines an equivalent SD gravity. In order for this expression to be well-defined, we have to additionally demand $\forall x\!\!:G(k(\phi(x)))>0$, which we have already encountered as a consistency requirement in Sec.~\ref{sssec:SDtoSTT}. 

On the other hand, we could also try to define $k(\phi)$ in such a way as to satisfy Eq.~(\ref{kinetic2}) for an arbitrary function $B(k(\phi))$. However, since $B(k(\phi))$ also depends on $k$, this would be a non-trivial differential equation, for which no general solution can be given. Therefore, we will limit our analysis to the simplest special cases. In the case $B(k(\phi))=0$, the term proportional to $\left( \frac{\partial k(\phi)}{\partial \phi} \right)^2$ on the left-hand side of Eq.~(\ref{kinetic2}) drops out and the STT must be equivalent to $f(R)$-gravity, as discussed in Sec.~\ref{sssec:SDtoSTT}. If $B$ is instead a non-zero constant $B_0$, we obtain
\begin{eqnarray}
k(\phi)
&=&
\pm \int d \phi \left[ \frac{1}{2B_0 A^2(\phi)}-\frac{3M_P^2}{B_0A^4(\phi)}\left( \frac{\partial A(\phi)}{\partial \phi} \right)^2 \right]^{1/2}~,
\end{eqnarray}
which is a smooth and well-defined solution for $k$ if $A^2$ is smooth and non-vanishing. Furthermore, in order for the argument of the square root to be positive we need $1-\frac{6M_P^2}{A^2(\phi)}\left( \frac{\partial A(\phi)}{\partial \phi} \right)^2 >0$.

In contrast to the situation encountered in Sec.~\ref{sssec:SDtoSTT}, it is not possible to uniquely determine an equivalent SD gravity theory for a given STT. In fact, we can only fix either $B(k)$ or $k$ in terms of the other. A convenient choice that resolves the problem of non-uniqueness is $k(\phi(x))=\phi(x)$, which is equivalent to identifying $k(x) \equiv \phi(x)$. After this identification, all quantities of the STT automatically depend on $x$ only through $k$, and Eqs.~(\ref{B3G}) and (\ref{B3Lambda}) are perfectly compatible with our assumptions about the structure of SD gravity, while Eq.~(\ref{Bsolution}) provides a unique expression
\begin{eqnarray}
    \label{B3B}
B(k(\phi)) = \left[ \frac{1}{2A^2(\phi)} - \frac{3M_P^2}{A^4(\phi)}\left( \frac{\partial A(\phi)}{\partial \phi} \right)^2 \right]~.
\end{eqnarray} 
We will use this particular choice when applying the equivalence to examples in Sec.~\ref{sec:Examples}. However, it is important to keep in mind that other equivalence relations are also possible. 


\subsection{Consistency check: mapping $f(R)$-gravity onto itself}
\label{sec:Consistency}

\begin{figure}[htbp]
\begin{center}
\includegraphics[scale=0.8]{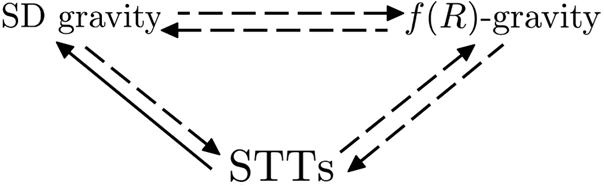}
\caption{
Triangle depicting the equivalence relations between SD gravity, $f(R)$-gravity and STTs; a solid line means that every model of the original type can be mapped to a theory of the resulting type, while a dotted line means that only specific models of the original type can consistently be mapped. 
Every SD theory with smooth couplings $G(k) >0$ and $B(k)$ fulfilling Eq.~(\ref{eq:third}) can be uniquely mapped onto a STT. In the other direction, every STT can be mapped onto an SD theory, however, not in a unique way. STTs can only be mapped to $f(R)$-gravity if their scale factors are of a particular form, see Eq.~(\ref{fRAphi}), and if Eq.~(\ref{eq:Rinvert}) is invertible. For the direction $f(R)$-gravity $\to$ STT, it is required that $f'(R)$ is invertible. Ref.~\cite{Calzada:2023yiq} states that only some SD theories can be mapped to $f(R)$-gravity, while $f(R)$-gravity is always equivalent to SD gravity. Though, the last is in contradiction with our findings, cp.~the mappings $f(R)$-gravity $\to$ SD gravity $\to$ STTs and $f(R)$-gravity $\to$ STTs, and we actually conclude from the equations in Ref.~\cite{Calzada:2023yiq} that the invertibility of $f'(R)$ is also required in the direction $f(R)$-gravity $\to$ SD gravity. Therefore, this figure depicts maps from $f(R)$-gravity to SD gravity with a dotted line.
}
\label{fig:EquiTri1}
\end{center}
\end{figure}

With the equivalence introduced in Sec.~\ref{sec:EquiAct} we have completed the equivalence triangle of SD gravity, $f(R)$-gravity and STTs; see Fig.~\ref{fig:EquiTri1}. Consequently, we can now perform an important consistency check by mapping $f(R)$-gravity onto itself along the triangle. It is known that $f(R)$-theory can be embedded into SD gravity \cite{Calzada:2023yiq} and into STTs \cite{Nojiri:2017ncd,Olmo:2006eh} if $f'(R) :=df/dR$ is invertible. However, only specific SD theories and STTs can be mapped to an equivalent $f(R)$-theory. 

For the consistency check, we will use the following Jordan frame action for $f(R)$-theory
\begin{eqnarray}
\label{SIIJ}
\Tilde{S}_{f(R)}&=&\frac{M_P^2}{2} \int d^4x \sqrt{-\Tilde{g}}f(\Tilde{R})~.
\end{eqnarray} 
Following Ref.~\cite{Calzada:2023yiq}, we can go from $f(R)$-gravity to a SD theory by defining a cosmological constant
\begin{eqnarray}
\label{fRLambda}
\Lambda(k(x))&=&\frac{1}{2}[\Tilde{R}(x)-M_P^2G(k(x))f(\Tilde{R}(x))]~,
\end{eqnarray} 
where $G(k(x))$ can be defined arbitrarily\footnote{As we have stated in Sec.~\ref{sssec:SDtoSTT}, a defining feature of SD gravity is that the running couplings depend on $x$ only through the scale-setting relation $k(x)$. However, this aspect of the equivalence was not considered in Ref.~\cite{Calzada:2023yiq}, i.e, it was not explained how $k(x)$ needs to be defined for Eq.~(\ref{fRLambda}) to be consistently satisfied. We expect that taking this issue into account will lead to additional constraints on either the resulting SD theories or the $f(R)$-theories, for example, the requirement for $f'(R)$ to be invertible. Those could enable us to find an explicit expression for $G(k)$, thus uniquely fixing the SD theory. Though, this is beyond the scope of the present paper, which is why we will assume that we can always find a suitable $k(x)$.}. As a result, the action in Eq.~(\ref{SIIJ}) can be written as 
\begin{eqnarray}
\frac{M_P^2}{2} \int d^4x \sqrt{-\Tilde{g}}f(\Tilde{R})
&\to&
\int d^4x \sqrt{-\Tilde{g}} \left( \frac{\Tilde{R} - 2\Lambda(k)}{2G(k)} \right)~,
\end{eqnarray} 
which coincides with the Jordan frame action of SD gravity for $B(k)=0$; see Eq.~(\ref{SIIIJ}). Note that $f(R)$-gravity will always lead to a SD theory with vanishing kinetic term. In the opposite direction, we can define $f(R)$-gravity from the quantities of the SD theory by
\begin{eqnarray}
\label{SDf}
f(\Tilde{R})&=&\frac{1}{M_P^2}\left( \frac{\Tilde{R} - 2\Lambda(k)}{G(k)} \right)~,
\end{eqnarray} 
which works only for SD gravities with $B(k)=0$. While it is generally necessary to know how to express the right-hand side of Eq.~(\ref{SDf}) in terms of $\Tilde{R}$ in order to complete the equivalence, this is not of concern for our present discussion since we start with $f(R)$-gravity when going along the equivalence triangle, such that all relevant quantities are always defined in terms of $\Tilde{R}$. 

Including STTs into the discussion, we can obtain a STT theory from $f(R)$-gravity via \cite{Kading:2019vyb,Olmo:2006eh}
\begin{eqnarray}
\label{fRA}
A^{-2}(\phi)&=&f'(\Tilde{R})~,
\\
\label{fRV}
V(\phi)&=&\frac{M_P^2}{2} \frac{1}{f'(\Tilde{R})^2} \left[ f'(\Tilde{R})\Tilde{R}-f(\Tilde{R}) \right]~,
\\
\label{fRphi}
\phi(x)&=&M_P \sqrt{\frac{3}{2}
} \ln f'(\Tilde{R})~.
\end{eqnarray} 
This results in a scale factor
\begin{eqnarray}
\label{fRAphi}
A^2(\phi)&=&e^{-\sqrt{\frac{2}{3}}\frac{1}{M_P}\phi
}~,
\end{eqnarray} 
which then leads to
\begin{eqnarray} 
\label{SIJk}
\Tilde{S}_\text{STT}&=&\int d^4x \sqrt{-\Tilde{g}}\left\{\frac{M_P^2}{2}\frac{\Tilde
R}{A^2(\phi)}-\frac{V(\phi)}{A^4(\phi)} \right\} 
\end{eqnarray} 
in agreement with the action in Eq.~(\ref{SIJ}) for $\kappa^2(\phi)=0$. When going into the opposite direction, i.e., from a STT to $f(R)$-gravity, we introduce $\Phi \coloneqq A^{-2}(\phi)$ and rewrite Eq.~(\ref{SIJk}) as
\begin{eqnarray}
\label{S1JkA}
\Tilde{S}_{\Phi}&=&\frac{M_P^2}{2}\int d^4x \sqrt{-\Tilde{g}}[\Phi \Tilde{R}-V(\Phi)] \, ,
\end{eqnarray} 
with
\begin{eqnarray}\label{SInewV}
V(\Phi)&=&\frac{2}{M_P^2}\frac{1}{A^4(\phi)}V(\phi)~.
\end{eqnarray} 
From Eq.~(\ref{S1JkA}) we find an equation of motion (EOM) for $\Phi$
\begin{eqnarray}
\label{eq:Rinvert}
\Tilde{R}&=&\frac{\partial V(\Phi)}{\partial \Phi}~,
\end{eqnarray} 
which needs to be invertible in order for us to find $\Phi(\Tilde{R})$ \cite{Olmo:2006eh}. We then obtain
\begin{eqnarray}
\label{fST}
f(\Tilde{R})=\Tilde{R}\Phi(\Tilde{R}) - V(\Phi(\Tilde{R}))~.
\end{eqnarray} 
Thus, for every STT with a scale factor of the form given in Eq.~(\ref{fRAphi}) and with $\frac{\partial V}{\partial \Phi}(\Tilde{R})$ being invertible, we get a unique equivalent $f(R)$-theory.

Having summarized all relevant equivalence relations, we can now map $f(R)$-gravity onto itself by following the equivalence triangle given in Fig.~\ref{fig:EquiTri1}. Starting from $f(R)$-gravity with an invertible $f'(\Tilde{R})$, we use Eq.~(\ref{fRLambda}) to define the cosmological constant of a SD theory with arbitrary $G(k)$ and $B(k)=0$. Using Eqs.~(\ref{A2aA1}) and (\ref{A2aV1}), we obtain a STT with $\kappa^2=0$ due to $B(k)=0$. In Sec.~\ref{sssec:SDtoSTT}, we have already seen that a STT of this type corresponds to $f(R)$-gravity. Since we find $V(\Phi)=\Tilde{R}\Phi(\Tilde{R})-f(\Tilde{R})$ with the same function $f(\tilde{R})$ we started with, this is consistent with Eqs.~(\ref{eq:Rinvert}) and (\ref{fST}), and it allows us to complete the map along the triangle from $f(R)$-gravity onto itself. Next, we can go along the triangle in the opposite direction. For this, we again start with $f(R)$-gravity and assume that $f'(\Tilde{R})$ is invertible. Using Eqs.~(\ref{fRA}) and (\ref{fRV}), we arrive at a STT with $\kappa^2(\phi)=0$. From Sec.~\ref{sec:STT2SD}, we know that STTs with this property are equivalent to SD theories with $B(k)=0$. In this case, we have no specific form of $k$, so any choice of $k(\phi(x))$ is applicable. Consequently, we can employ Eq.~(\ref{SDf}) in order to recover the action of $f(R)$-gravity with the same function $f(\tilde{R})$ we started with, meaning that we again went once around the triangle. This demonstrates the consistent embedding of the equivalence found in Sec.~\ref{sec:EquiAct} into the equivalence triangle in Fig.~\ref{fig:EquiTri1}.


\subsection{Equivalence on the level of field equations}
\label{sec:FieldEqs}

It is known that the equivalences between $f(R)$-gravity and SD gravity or STTs on the level of the actions automatically imply equivalences on the level of field equations \cite{Calzada:2023yiq,Olmo:2006eh}. Therefore, we will now investigate the equivalence between SD gravity and STTs on the level of the field equations as another consistency check. This also gives us the opportunity to further analyze the two classes of theories, and this will uncover an important relation between the EOM for $\phi$ and a specific scale-setting method used in SD gravity.

For SD gravity and STTs, we will need to derive the Jordan frame Einstein equations, which are modified not only due to the presence of the kinetic terms for $k$ and $\phi$, but also due to the non-minimal coupling of the Ricci scalar to $\frac{1}{G(k)}$ and $\frac{1}{A^2(\phi)}$. In addition, for STTs, we will have an EOM for $\phi$. Note that, for SD gravity, the action in Eq.~(\ref{SIIIJ}) is not varied with respect to $G$ and $\Lambda$ because the running couplings are not obtained from a variational principle but rather from solving the renormalization group equations once a scale-setting relation $k(x)$ has been fixed \cite{Reuter:2003ca}. The couplings can then be seen as externally prescribed background fields in the context of the variational principle. Since, as of yet, there is no unique way of fixing $k$ and various methods are discussed in the literature~\cite{Bertini:2024naa,Koch:2014joa,Contreras:2016mdt,Domazet:2012tw}, the focus is often not on finding $k(x)$ itself, but rather on determining the functions $G(k(x))$ and $\Lambda(k(x))$. One method is the so-called effective Lagrangian fixed point condition (ELFP)~\cite{Koch:2014joa,Contreras:2016mdt}, which reads
\begin{eqnarray} 
\label{ELFP}
\frac{\partial}{\partial k}\Tilde{S}_\text{SD} &=& 0~.
\end{eqnarray}
For $B(k)=0$, this is nothing but the field equation of $k$ obtained from the variational principle. An EOM for $k$ can thus be seen as an appropriate generalization of this condition to the situation where we have a kinetic term for $k$ in the action.

In Appendix \ref{app:ModEinsEqs}, we derive the modified Einstein equations for SD gravity
\begin{eqnarray} 
\label{EinsteinSIIIJ}
\Tilde{G}_{\mu \nu}+\Lambda(k)\Tilde{g}_{\mu \nu} &=& -\Delta_{G}t_{\mu \nu} - \Delta_{B}t_{\mu \nu}~,
\end{eqnarray}
where $\Tilde{G}_{\mu \nu}$ is the Einstein tensor, and the contributions to the effective energy-momentum tensor coming from the scale-dependence of the Newton constant and the newly added kinetic term for $k$ are given by
\begin{eqnarray} 
\label{DtG}
\Delta_{G}t_{\mu \nu} 
&:=&
\frac{1}{G(k)} \left[ \Tilde{\nabla}_{\nu} \Tilde{\nabla}_{\mu} G(k) - \frac{2}{G(k)} \Tilde{\nabla}_{\mu} G(k) \Tilde{\nabla}_{\nu} G(k) + \frac{2}{G(k)}\Tilde{g}_{\mu \nu}(\Tilde{\nabla}G(k))^2 - \Tilde{g}_{\mu \nu}\Tilde{\Box}G(k) \right] ~,~~~
\end{eqnarray}
and
\begin{eqnarray}  
\label{DtB}
\Delta_{B}t_{\mu \nu} 
&:=&
G(k)B(k)\left[\Tilde{g}_{\mu \nu}(\Tilde{\partial}k)^2-2(\Tilde{\partial}_{\mu}k)(\Tilde{\partial}_{\nu}k)\right]~.
\end{eqnarray}
For future uses, we express Eq.~(\ref{DtG}) in terms of derivatives of $k$
\begin{eqnarray} 
\label{DtGk}
\Delta_{G}t_{\mu \nu} &=&\frac{1}{G(k)}\left( \frac{\partial G(k)}{\partial k} \right) \left[ \Tilde{\nabla}_{\nu} \Tilde{\nabla}_{\mu} k - \Tilde{g}_{\mu \nu} \Tilde{\Box}k \right] 
\nonumber
\\ 
&&
+\frac{1}{G(k)} \left( \frac{2}{G(k)} \left( \frac{\partial G(k)}{\partial k} \right)^2 -\frac{\partial^2 G(k)}{\partial k^2} \right) \left[\Tilde{g}_{\mu \nu}(\Tilde{\partial}k)^2 - (\Tilde{\partial}_{\mu}k)(\Tilde{\partial}_{\nu}k) \right]~.
\end{eqnarray}
Analogously, for STTs, we find
\begin{eqnarray} 
\label{EinsteinSIJ}
\Tilde{G}_{\mu \nu}+\frac{V(\phi)}{M_P^2A^2(\phi)}\Tilde{g}_{\mu \nu} &=& -\Delta_{A}t_{\mu \nu} - \Delta_{\kappa}t_{\mu \nu} ~,
\end{eqnarray}
with
\begin{eqnarray} 
\label{DtA}
\Delta_{A}t_{\mu \nu} 
&:=&
\frac{2}{A(\phi)}\left( \frac{\partial A(\phi)}{\partial \phi} \right) \left[ \Tilde{\nabla}_{\nu} \Tilde{\nabla}_{\mu} \phi - \Tilde{g}_{\mu \nu} \Tilde{\Box}\phi \right] 
\nonumber
\\ 
&&
+\frac{2}{A(\phi)} \left( \frac{3}{A(\phi)} \left( \frac{\partial A(\phi)}{\partial \phi} \right)^2 -\frac{\partial^2 A(\phi)}{\partial \phi^2} \right) \left[\Tilde{g}_{\mu \nu}(\Tilde{\partial}\phi)^2 - (\Tilde{\partial}_{\mu}\phi)(\Tilde{\partial}_{\nu}\phi) \right] \, ,
\end{eqnarray}
and
\begin{eqnarray}  
\label{Dtkappa}
\Delta_{\kappa}t_{\mu \nu} 
&:=&
\frac{1}{2M_P^2}\Tilde{g}_{\mu \nu}\kappa^2(\phi) (\Tilde{\partial}\phi)^2 - \frac{1}{M_P^2}\kappa^2(\phi) (\Tilde{\partial}_{\mu}\phi)(\Tilde{\partial}_{\nu}\phi)
~.
\end{eqnarray}

Next, we have a look at the EOMs for $\phi$ and $k$ in the Jordan frame. Using the Euler-Lagrange equations, we find for $\phi$
\begin{eqnarray}  
\label{EOMphi}
&&
-M_P^2\Tilde{R}\frac{1}{A(\phi)^3}\frac{\partial A(\phi)}{\partial \phi} + 4V(\phi)\frac{1}{A(\phi)^5}\frac{\partial A(\phi)}{\partial \phi} -\frac{\partial V(\phi)}{\partial \phi}\frac{1}{A^4(\phi)} + \kappa^2(\phi) \Tilde{\Box}\phi \frac{1}{A^2(\phi)} 
\nonumber
\\
&&
+ (\Tilde{\partial}\phi)^2 \frac{1}{A^3(\phi)}\frac{\partial A(\phi)}{\partial \phi} \left[ -\kappa^2(\phi)+6M_P^2 \frac{1}{A^2(\phi)} \left( \frac{\partial A(\phi)}{\partial \phi} \right)^2 -6M_P^2 \frac{1}{A(\phi)} \frac{\partial^2 A(\phi)}{\partial \phi^2} \right]=0 \, .
\end{eqnarray}
We see that Eq.~(\ref{EOMphi}) contains the Ricci scalar, which effectively couples the EOM for $\phi$ to the Einstein equations and makes finding solutions for it an intricate endeavor. Again, using the Euler-Lagrange equations, we obtain
\begin{eqnarray}
\label{EOMk}
-\Tilde{R}\frac{1}{2G(k)^2}\frac{\partial G}{\partial k} + \frac{1}{G(k)^2}\frac{\partial G}{\partial k} \Lambda(k) - \frac{1}{G(k)} \frac{\partial \Lambda(k)}{\partial k} + 2B(k)\Tilde{\Box}k +\frac{\partial B(k)}{\partial k} (\Tilde{\partial}k)^2 &=& 0~,
\end{eqnarray}
as the EOM for $k$.

Before actually checking the equivalence on the level of the field equations, we will now briefly discuss the divergence of the modified energy-momentum tensor. Even though the Einstein equations are modified in SD gravity and STTs, the Bianchi identities still ensure that the divergence of the Einstein tensor vanishes. Therefore, for SD gravity, 
\begin{eqnarray} 
\label{olddiv}
0 &=& \Tilde{\nabla}^{\mu}\Tilde{G}_{\mu \nu} = -\Tilde{\nabla}^{\mu} \left[ \Lambda(k)\Tilde{g}_{\mu \nu} + \Delta_{G}t_{\mu \nu} + \Delta_{B}t_{\mu \nu} \right] \, ,
\end{eqnarray}
should hold and there is a similar relation for STTs. Note that we only demand that Eq.~(\ref{olddiv}) is satisfied on-shell, i.e., up to terms proportional to the EOM. In STTs, we have the EOM for $\phi$ at our disposal, while, in a SD theory, the EOM for $k$ is not necessarily imposed. It turns out that in a theory with a non-minimal coupling of the Ricci scalar to a function of $x$, Eq.~(\ref{olddiv}) makes it difficult to make contact with the EOM (\ref{EOMk}). The reason for this is that when extracting the Einstein equations (\ref{EinsteinSIIIJ}) from the variation of the action, we usually pull out a factor $\frac{1}{2G(k)}$, see Eq.~(\ref{varSIIIJfull}), which does depend on $x$. We find that it is much simpler to make contact with the EOM by explicitly including this factor in the Einstein equations, which means working with
\begin{eqnarray} 
\label{newdiv}
0 &=& \frac{1}{2G(k)}\nabla^{\mu}\Tilde{G}_{\mu \nu} = -\Tilde{\nabla}^{\mu} \left[\frac{1}{2G(k)} \Lambda(k)\Tilde{g}_{\mu \nu} + \frac{1}{2G(k)}\Delta_{G}t_{\mu \nu} + \frac{1}{2G(k)}\Delta_{B}t_{\mu \nu}   \right] -\Tilde{G}_{\mu \nu} \Tilde{\nabla}^{\mu}\left( \frac{1}{2G(k)} \right)~.
\nonumber
\\
\end{eqnarray}
In what follows, we will show that Eq.~(\ref{newdiv}) is indeed satisfied provided that we impose a suitable condition on $k(x)$. Substituting the explicit expressions from Appendix \ref{app:Divergence} into Eq.~(\ref{newdiv}) gives
\begin{eqnarray} 
2 \left[ -\Tilde{R}\frac{1}{2G(k)^2}\frac{\partial G}{\partial k} + \frac{1}{G(k)^2}\frac{\partial G}{\partial k} \Lambda(k) - \frac{1}{G(k)} \frac{\partial \Lambda(k)}{\partial k} + 2B(k)\Tilde{\Box}k +\frac{\partial B(k)}{\partial k} (\Tilde{\partial}k)^2 \right]\Tilde{\nabla}_{\nu}k &=&0~,~
\end{eqnarray}
which coincides with the EOM for $k$ (\ref{EOMk}) multiplied by $2\nabla_{\nu}k$. Thus, imposing Eq.~(\ref{EOMk}) is a sufficient condition for the vanishing of the divergence of the effective energy-momentum tensor. Note that working with Eq.~(\ref{newdiv}) instead of Eq.~(\ref{olddiv}) is merely a matter of preference since we can arrive at the same conclusion by working with either of them. Performing a similar calculation for STTs yields the same result, i.e., the appropriate analogue of Eq.~(\ref{newdiv}) is proportional to the EOM for $\phi$ (\ref{EOMphi}) and thus vanishes on-shell. So, while STTs are always anomaly-free on-shell, for SD gravity we need the additional constraint (\ref{EOMk}). In particular, note that Eq.~(\ref{EOMk}) is valid only for the specific action considered here. For effective actions that include additional ingredients, such as an electromagnetic source, for instance, the condition Eq.~(\ref{EOMk}) must be generalized to incorporate the running of the electromagnetic coupling function. Ref.~\cite{Reuter:2003ca} analyzed the divergence of the effective energy-momentum tensor of SD gravity with an additional contribution to the energy-momentum tensor describing the four-momentum carried by the running couplings. Since Ref.~\cite{Reuter:2003ca} did not treat $k$ as a dynamical field, no explicit kinetic term for $k$ was included in the action and no EOM for $k$ was imposed. Instead, in order to guarantee the vanishing divergence of the energy-momentum tensor, suitable choices of the additional contribution to the energy-momentum tensor, the matter-part of the energy-momentum tensor, as well as the cosmological constant, were discussed. One particular example was 
$G(k)\Lambda(k)\!\!: \text{const.}$ and
$T_{\mu \nu}=0$, where $T_{\mu \nu}$ is the matter part of the energy-momentum tensor. In this case, it was found that the additional contribution has to be
\begin{eqnarray} 
\theta_{\mu \nu} &=& \frac{3}{4}\frac{1}{G(k)^2}\left( \frac{\partial G(k)}{\partial k} \right)^2 [\Tilde{g}_{\mu \nu}(\Tilde{\partial} k)^2 - 2 (\Tilde{\nabla}_{\mu}k)(\Tilde{\nabla}_{\nu}k)]~,
\end{eqnarray}
which is very similar to the contribution in Eq.~(\ref{DtB}) arising from the kinetic term for $k$. More precisely, if we make the particular choice
\begin{eqnarray} 
\label{Bspec}
B &=& \frac{3}{4}\frac{1}{G(k)^3}\left( \frac{\partial G(k)}{\partial k} \right)^2~,
\end{eqnarray}
we obtain $\theta_{\mu \nu} = \Delta_{B}t_{\mu \nu}$. Moreover, Ref.~\cite{Reuter:2003ca} found that $\theta_{\mu \nu}$ can arise from a term in the action
\begin{eqnarray} 
\label{eq:ReuterAction}
\Tilde{S}_{\theta} &\propto& \int d^4x \sqrt{-\Tilde{g}} e^{\psi} (\Tilde{\partial} \psi)^2~,
\end{eqnarray}
where $\psi = -\ln(G)$. Rewriting Eq.~(\ref{eq:ReuterAction}) in terms of $G(k)$, we get
\begin{eqnarray} 
\Tilde{S}_{\theta} &\propto& \int d^4x \sqrt{-\Tilde{g}} \frac{1}{G(k)^3} \left( \frac{\partial G(k)}{\partial k} \right)^2 (\Tilde{\partial}k)^2~,
\end{eqnarray}
which, up to a numerical prefactor that we can leave out, coincides with the kinetic term in Eq.~(\ref{SIIIJ}) for the choice of $B(k)$ made in Eq.~(\ref{Bspec}). Thus, we have found a kinetic term for $k$, such that the situation in Ref.~\cite{Reuter:2003ca} can be treated as a special case of our discussion for the divergence of the energy-momentum tensor corresponding to the action in Eq.~(\ref{SIIIJ}), in which the particular choices (\ref{Bspec}) and $G(k)\Lambda(k)\!\!: \text{const.}$ lead to a vanishing divergence of the energy-momentum tensor without the need to impose the EOM for $k$.

Finally, we will now study the equivalence of the field equations. When starting with SD gravity, we use Eqs.~(\ref{A2aA1}), (\ref{A2aV1}) and (\ref{kin1}) as well as
\begin{eqnarray}
\label{t3}
\frac{\partial \phi}{\partial k} \Tilde{\nabla}_{\nu} \Tilde{\nabla}_{\mu}k &=& \Tilde{\nabla}_{\nu} \Tilde{\nabla}_{\mu}\phi - \frac{\partial^2 \phi}{\partial k^2} (\Tilde{\nabla}_{\nu}k) (\Tilde{\nabla}_{\mu}k)~,~~~~
\frac{\partial \phi}{\partial k} \Tilde{\Box}k = \Tilde{\Box}\phi - \frac{\partial^2 \phi}{\partial k^2} (\Tilde{\partial}k)^2~,
\end{eqnarray} 
such that we find for the individual terms in the modified Einstein equations (\ref{EinsteinSIIIJ})
\begin{eqnarray}
\Lambda(k)\Tilde{g}_{\mu \nu} &=& \frac{V(\phi(k))}{M_P^2A^2(\phi(k))}\Tilde{g}_{\mu \nu}~,
\\
\Delta_{B}t_{\mu \nu} 
&=& \frac{1}{2M_P^2}\kappa^2(\phi) \left( \frac{\partial \phi(k)}{\partial k} \right)^2 \left[\Tilde{g}_{\mu \nu}(\Tilde{\partial}k)^2-2(\Tilde{\partial}_{\mu}k)(\Tilde{\partial}_{\nu}k)\right] 
= \Delta_{\kappa}t_{\mu \nu}~,
\end{eqnarray} 
and
\begin{eqnarray}
\label{long1}
\Delta_{G}t_{\mu \nu} 
&=&\frac{2}{A(\phi)} \frac{\partial A(\phi)}{\partial \phi} \left[ \Tilde{\nabla}_{\nu} \Tilde{\nabla}_{\mu} \phi - \frac{\partial^2 \phi}{\partial k^2}(\Tilde{\partial}_{\mu}k)(\Tilde{\partial}_{\nu}k) - \Tilde{g}_{\mu \nu} \Tilde{\Box}\phi + \Tilde{g}_{\mu \nu} \frac{\partial^2 \phi}{\partial k^2} (\Tilde{\partial}k)^2 \right] 
\nonumber
\\
&&+ \Bigg( \frac{8}{A^2(\phi)}\left( \frac{\partial A(\phi)}{\partial \phi} \right)^2 \left( \frac{\partial \phi}{\partial k} \right)^2
- 2\frac{1}{A(\phi)} \frac{\partial A(\phi)}{\partial \phi} \frac{\partial^2 \phi}{\partial k^2} 
\nonumber
\\
&&- \frac{2}{A^2(\phi)} \left( \frac{\partial A(\phi)}{\partial \phi} \right)^2 \left( \frac{\partial \phi}{\partial k} \right)^2 - \frac{2}{A(\phi)} \frac{\partial^2 A(\phi)}{\partial \phi^2} \left( \frac{\partial \phi}{\partial k} \right)^2 \Bigg)\left[\Tilde{g}_{\mu \nu}(\Tilde{\partial}k)^2 - (\Tilde{\partial}_{\mu}k)(\Tilde{\partial}_{\nu}k) \right]  
\nonumber
\\
&=&
 \Delta_{A}t_{\mu \nu}~,
\end{eqnarray} 
which shows that this direction of the equivalence is working for the Einstein equations. Similarly, starting from Eq.~(\ref{EOMk}), we find
\begin{eqnarray}
&& \frac{\partial \phi}{\partial k}\Bigg[ -M_P^2\Tilde{R}\frac{1}{A(\phi)^3}\frac{\partial A(\phi)}{\partial \phi} + 4V(\phi)\frac{1}{A(\phi)^5}\frac{\partial A(\phi)}{\partial \phi} -\frac{\partial V(\phi)}{\partial \phi}\frac{1}{A^4(\phi)} + \kappa^2(\phi) \Tilde{\Box}\phi \frac{1}{A^2(\phi)} 
\nonumber
\\
&&+(\Tilde{\partial}\phi)^2 \frac{1}{A^3(\phi)}\frac{\partial A(\phi)}{\partial \phi} \left( -\kappa^2(\phi)+6M_P^2 \frac{1}{A^2(\phi)} \left( \frac{\partial A(\phi)}{\partial \phi} \right)^2 -6M_P^2 \frac{1}{A(\phi)} \frac{\partial^2 A(\phi)}{\partial \phi^2} \right) \Bigg] =0~,
\end{eqnarray} 
which is nothing but Eq.~(\ref{EOMphi}) multiplied by $\frac{\partial \phi}{\partial k}$. From this, we can conclude that if we impose the EOM for $k$, any $\phi(k)$ will satisfy its own EOM as well. On the other hand, if we start with a SD theory in which Eq.~(\ref{EOMk}) is not imposed, there is no guarantee that $\phi(k)$ given by Eq.~(\ref{A2a3}) will satisfy Eq.~(\ref{EOMphi}). For the other direction of the equivalence, we need Eqs.~(\ref{B3G}), (\ref{B3Lambda}) and (\ref{kinetic2}), such that we obtain for the individual terms in the modified Einstein equations (\ref{EinsteinSIJ})
\begin{eqnarray} 
\frac{V(\phi)}{M_P^2A^2(\phi)}\Tilde{g}_{\mu \nu} &=& \Lambda(k)\Tilde{g}_{\mu \nu}~,
\\
\Delta_{\kappa}t_{\mu \nu} 
&=& G(k)B(k)\Tilde{g}_{\mu \nu} \left( \frac{\partial k}{\partial \phi} \right)^2 (\Tilde{\partial}\phi)^2 - 2G(k)B(k) \left( \frac{\partial k}{\partial \phi} \right)^2 (\Tilde{\partial}_{\mu}\phi)(\Tilde{\partial}_{\nu}\phi)  = \Delta_{B}t_{\mu \nu}~,
\end{eqnarray} 
and
\begin{eqnarray} 
\Delta_{A}t_{\mu \nu} 
&=&  \frac{1}{G(k)}\frac{\partial G(k)}{\partial k} \left[ \nabla_{\nu} \nabla_{\mu} k - \frac{\partial^2 k}{\partial \phi^2}(\nabla_{\mu}\phi)(\nabla_{\nu}\phi) - \Tilde{g}_{\mu \nu} \Tilde{\Box}k + \frac{\partial^2 k}{\partial \phi^2} \Tilde{g}_{\mu \nu} (\Tilde{\partial}\phi)^2 \right] 
\nonumber
\\
&&+ \Bigg( \frac{1}{G(k)}\frac{\partial G(k)}{\partial k}\frac{\partial^2 k}{\partial \phi^2}+\frac{1}{G(k)} \frac{\partial^2 G(k)}{\partial k^2}\left( \frac{\partial k}{\partial \phi} \right)^2 
\nonumber
\\
&&- \frac{1}{2G(k)^2} \left( \frac{\partial G(k)}{\partial k} \right)^2 \left( \frac{\partial k}{\partial \phi} \right)^2 - \frac{3}{2G(k)^2} \left( \frac{\partial G(k)}{\partial k} \right)^2 \left( \frac{\partial k}{\partial \phi} \right)^2  \Bigg)\left[(\nabla_{\mu}\phi)(\nabla_{\nu}\phi) -\Tilde{g}_{\mu \nu}(\Tilde{\partial}\phi)^2 \right] 
\nonumber
\\
 &=& \Delta_{G}t_{\mu \nu}~,
\end{eqnarray} 
which shows that the equivalence for the Einstein equations holds. In addition, starting from Eq.~(\ref{EOMphi}), the equivalence gives
\begin{eqnarray} 
\frac{\partial k}{\partial \phi}\left[ -\Tilde{R}\frac{1}{2G(k)^2}\frac{\partial G}{\partial k} + \frac{1}{G(k)^2}\frac{\partial G}{\partial k} \Lambda(k) - \frac{1}{G(k)} \frac{\partial \Lambda(k)}{\partial k} + 2B(k)\Tilde{\Box}k +\frac{\partial B(k)}{\partial k} (\Tilde{\partial}k)^2 \right] &=&0~,
\end{eqnarray} 
which coincides with Eq.~(\ref{EOMk}) multiplied by $\frac{\partial k}{\partial \phi}$. This means that in a SD theory obtained from a STT, $k$ will always satisfy its EOM. 

In summary, we have seen that the field equations of one theory can be mapped onto the field equations of the other without the need for further restrictions. A consistent STT will always result in a SD theory in which $k$ satisfies its EOM. On the other hand, it is necessary to impose Eq.~(\ref{EOMk}) in a SD theory in order to arrive at a STT in which $\phi$ satisfies Eq.~(\ref{EOMphi}). The same condition also guarantees that the SD theory is anomaly-free, i.e., that the divergence of the effective energy-momentum tensor vanishes, as required by the Bianchi identities. This intimate relationship between the EOMs for $k$ and $\phi$ leads to the conclusion that only SD theories in which the ELFP-condition is imposed are relevant for the equivalence between SD gravity and STTs.


\subsection{Equivalence in the Einstein frame}
\label{sec:Einstein}

In Sec.~\ref{sec:FieldEqs}, we have seen that the EOMs of SD gravity and STTs in the Jordan frame are rather complicated and coupled to the Einstein equations, which renders the search for solutions intricate. However, when dealing with concrete models, we want to give explicit solutions for the EOMs. From STTs we know that the EOMs in the Einstein frame are much simpler than in the Jordan frame; see, e.g., Ref.~\cite{MarioHabil}. Furthermore, there are no couplings to the Einstein equations in this frame. Therefore, we will now briefly discuss the Einstein frame version of the equivalence between SD gravity and STTs. 

The STT action in the Einstein frame is given by Eq.~(\ref{SI}), and in order to find an action for SD gravity in the same frame, we need to apply a conformal transformation to the Jordan frame action in Eq.~(\ref{SIIIJ}), i.e., using Eq.~(\ref{Aequiv}), we have  
\begin{eqnarray}
\label{eq:conftrafoSD}
\Tilde{g}_{\mu \nu} &=& M_P^2G(k) g_{\mu \nu}~.
\end{eqnarray}
We also need the inverse transformation law for the Ricci scalar given by
\begin{eqnarray}
\Tilde{R} &=& \frac{1}{M_P^2G(k)}\left( R - 6 \Box \ln (M_P \sqrt{G(k)}) - 6(\partial \ln (M_P \sqrt{G(k)}))^2 \right)~.
\end{eqnarray}
As a consequence, the Einstein frame action of SD gravity is obtained
\begin{eqnarray}
\label{SIII}
S_\text{SD} &=& \int d^4x \sqrt{-g} \left[ \frac{M_P^2}{2}R - \Bar{B}(k)(\partial k)^2 - M_P^4G(k)\Lambda(k) \right]~,
\end{eqnarray}
where
\begin{eqnarray}
\label{BBar}
\Bar{B}(k) &:=& M_P^2 \left( G(k)B(k) + \frac{3}{4}\left( \frac{\partial \ln G(k)}{\partial k} \right)^2 \right)~.
\end{eqnarray}
Comparing Eq.~(\ref{SIII}) to Eq.~(\ref{SI}), we recover Eq.~(\ref{A2aV1}) and we find
\begin{eqnarray}
\label{EA2aB}
\Bar{B}(k)=\frac{1}{2} \left( \frac{\partial \phi}{\partial k} \right)^2~.
\end{eqnarray}
Substituting Eq.~(\ref{BBar}) into Eq.~(\ref{EA2aB}), we arrive at Eq.~(\ref{ODE}). Since we can simply apply the identification in Eq.~(\ref{A2aA1}), we now have fully re-derived the equivalence relations from Sec.~\ref{sssec:SDtoSTT} that map SD gravity to a STT. Completely analogously, we could also again derive the relations given in Sec.~\ref{sec:STT2SD}.

The Einstein equations in the Einstein frame can be calculated in a straightforward way due to the absence of a non-minimal coupling to the Ricci scalar. For STTs, we obtain
\begin{eqnarray} 
\label{EinsteinSTE}
G_{\mu \nu} + \frac{V(\phi)}{M_P^2}g_{\mu \nu} &=& -\Delta_{\phi}t_{\mu \nu}~,
\end{eqnarray}
where
\begin{eqnarray} 
\label{Dtphi}
\Delta_{\phi}t_{\mu \nu} 
&:=&
-\frac{1}{M_P^2}(\nabla_{\mu}\phi)(\nabla_{\nu}\phi) + \frac{1}{2M_P^2} (\partial \phi)^2 g_{\mu \nu}~,
\end{eqnarray}
which is a contribution to the effective energy-momentum tensor and arises from the kinetic term of $\phi$ analogously to $\Delta_{\kappa}t_{\mu \nu}$ in the Jordan frame; see Eq.~(\ref{Dtkappa}). For SD gravity, the Einstein equations are
\begin{eqnarray} 
G_{\mu \nu} + M_P^2 G(k)\Lambda(k)g_{\mu \nu} &=& - \Delta_{k}t_{\mu \nu} ~,
\end{eqnarray}
with
\begin{eqnarray}
\Delta_{k}t_{\mu \nu} &:=& \frac{1}{M_P^2}\Bar{B}(k) (\partial k)^2 g_{\mu \nu} - \frac{2}{M_P^2}\Bar{B}(k) (\nabla_{\mu}k)(\nabla_{\nu}k)~.
\end{eqnarray}
By using the Euler-Lagrange equations, we also obtain the EOMs for $\phi$ and for $k$
\begin{eqnarray}
\label{EOMphiE}
\Box \phi - \frac{\partial V(\phi)}{\partial \phi} &=& 0~,
\\
\label{EOMkE}
2 \Bar{B}(k)\Box k + \frac{\partial \Bar{B}(k)}{\partial k}(\partial k)^2 - M_P^4\frac{\partial}{\partial k}(G(k)\Lambda(k)) &=& 0~.
\end{eqnarray}
Given Eq.~(\ref{BBar}), the EOM (\ref{EOMkE}) appears to be rather complicated. However, it can be simplified when considering a SD theory with $B(k)=0$, which applies to essentially all models in the literature \cite{Rincon:2019ptp,Alvarez:2022wef,Alvarez:2022mlf,Alvarez:2023ywi,Rincon:2022hpy}. Furthermore, in most of these models, it is not required that $k$ satisfies its EOM. In such cases, we only need to impose the EOM for $\phi$. However, as we have seen in Sec.~\ref{sec:FieldEqs}, it might not always be possible for $\phi$ to satisfy its EOM if the one for $k$ is not imposed. Another option would be to start from a STT and make the choice $k(\phi)=\phi$; see Sec.~\ref{sec:STT2SD}. In this case, Eq.~(\ref{EA2aB}) reduces to $\Bar{B}(k) = 1/2$, such that Eq.~(\ref{EOMkE}) becomes
\begin{eqnarray} 
\label{B3EOMkE}
\Box k - M_P^4\frac{\partial}{\partial k}(G(k)\Lambda(k)) &=& 0~,
\end{eqnarray}
which is almost as simple as the EOM for $\phi$. 

Finally, it should be stated that also for Einstein frame field equations we can perform the same mapping as for the Jordan frame in Sec.~\ref{sec:FieldEqs}, which is another passed consistency check of the equivalence between SD gravity and STTs in the Einstein frame. In addition, for both theories, the divergence of the effective energy-momentum tensor still vanishes after employing the respective EOM. The fact that all aspects of the equivalence are unchanged under conformal transformations now allows us to freely switch between the Einstein frame and the Jordan frame.


\section{Examples of equivalence pairs}
\label{sec:Examples}

In this section, we will study examples of equivalence pairs. At first, we will briefly discuss consistency conditions that we can use to assess the sanity of physical models. Subsequently, we will apply the equivalence relations to some of the most discussed STTs and SD models.


\subsection{Consistency conditions for physical models}

Several physically motivated conditions are usually imposed on a model to ensure that it is consistent with observations. For STTs, the first condition that we want to discuss is the existence of a lower bound of the scalar potential. By demanding
\begin{eqnarray}
\forall x\!\!:V(\phi(x)) &\geq& -c \, ,
\end{eqnarray}
with $c$ being a finite constant, we avoid the existence of infinite negative energy states, as required by the observed stability of matter. Since Eq.~(\ref{A2aV1}) dictates that $V(\phi) \sim G(k)\Lambda(k)$ and we require $G(k)>0$, this implies that only SD theories for which $\Lambda$ is bounded from below can result in physically acceptable STTs.

Other important aspects are the signs of the kinetic terms for $\phi$ and $k$. It is known that a kinetic term with the wrong sign in the action leads to ghost-like instabilities of the corresponding theory \cite{Delhom:2022vae}. From Eq.~(\ref{Bequiv}), we can immediately see that for both directions of the equivalence the kinetic term of the resulting theory will have the same sign as the original one. Thus, starting with a theory that is free of instabilities, this property will be inherited by the equivalent theory.

Next, we will discuss fall-off conditions for $\phi$ and $k(x)$. In order to safely drop boundary terms from the action, it must be assumed that all fields fall off sufficiently fast at infinity and settle to a constant value. For the scalar field solutions that satisfy such fall-off conditions, and for our standard choice of $k(x)=\phi(x)$, these properties are inherited by the scale-setting relation. Though, attention must be paid when working with a more complicated $k(\phi)$. The situation is more complicated when starting from a SD theory. In principle, if $k(x)$ and the running couplings of the theory are known, Eq.~(\ref{A2a3}) gives the resulting behavior of $\phi(k)$. However, in the SD literature it is customary to bypass the definition of $k(x)$ by directly determining the running couplings as functions of $x$. 
The rationale for this selection is well-founded: the adoption of the scale \( k \) as a function of the coordinates, such as \( k \rightarrow k(x^{\mu}) \), generally imposes limitations on the predictability of the derived physical quantities. Consequently, it is prudent to mitigate potential anomalies by avoiding the specification of a particular functional form for \( k(x^{\mu}) \), which is the common approach in the majority of cases with $B(k)=0$.
While we can still draw conclusions from Eq.~(\ref{A2a3}), it will, in general, not be possible to evaluate the integral over $k$ explicitly. Though, in the usually employed case with $B(k)=0$, the integral can be evaluated without any information about the specific form of $k(x)$.

In addition, of relevance are energy conditions such as the null-energy-condition (NEC) 
\begin{eqnarray}
    \label{NEC}
    \forall l^{\mu} \in \{ a^\rho| a^{\rho}a_{\rho}=0 \}\!\!:
T^{\text{eff}}_{\mu \nu}l^{\mu}l^{\nu} \geq 0
~,
\end{eqnarray}
where $T^{\text{eff}}_{\mu \nu}$ is the sum of all contributions to the energy-momentum tensor arising in the modified Einstein equations. We will later see that the NEC can be used in SD gravity to fix the running couplings of the theory. The same can be done by employing the ELFP, which, however, leads to a different theory. From this, we can conclude that imposing both conditions at the same time might overconstrain the theory. Recalling the important role of the ELFP for the equivalence, it would certainly be interesting to investigate whether there exist theories for which both conditions can simultaneously be satisfied. We leave this subject for future work.


\subsection{STTs as SD theories}
\label{sec:ExSTTasSD}

Here, we will apply the equivalence to three popular STTs: conformally coupling chameleons, symmetrons and environment-dependent dilatons. For this, we will follow the presentation in Ref.~\cite{Fischer:2024eic}, and we will work in the Einstein frame. In the Einstein frame, the conformal coupling to the metric tensor induces an effective coupling to matter, see, for example, Ref.~\cite{Kading:2019vyb}, such that the scalar $\phi$ has an effective potential
\begin{eqnarray}
V_{\text{eff}}(\phi) &=& V(\phi) + \rho(x)A(\phi)~,
\end{eqnarray}
where $\rho(x)$ is the density of non-relativistic matter. The matter coupling is expected to cause a fifth force, which, however, is screened in dense environments due to the interplay with $V(\phi)$, such that tight Solar System-based constraints on fifth forces \cite{Dickey1994,Adelberger2003,Kapner2007} can be circumvented. Since, in Sec.~\ref{sec:Einstein}, we have shown that SD gravity can also be expressed in the Einstein frame, we would expect that the conformal coupling of the form given in Eq.~(\ref{eq:conftrafoSD}) will also lead to an additional coupling to matter, whose significance could be explored in future works. The scalar potential and scale factor that define each STT are listed for the three considered models in Tab.~\ref{tab:models}.
\begin{table}[htbp]
\centering
\renewcommand{\arraystretch}{1.5}
\begin{tabular}{|c||c|c|}
\hline\
Scalar Field & $V(\phi)$ & $A(\phi)$ \\
\hline\hline
\textit{Chameleon} & $\displaystyle\frac{\Lambda^{n+4}}{\phi^n}$ &$\displaystyle{e}^{\phi / M_C}$ \\
\hline
\textit{Symmetron}  & $\displaystyle-\frac{\mu^2}{2}\,\phi^2 + \frac{\lambda_S}{4}\,\phi^4$ & $\displaystyle1 + \frac{\phi^2}{2M_S^2}$ \\
\hline
\textit{Dilaton} & $\displaystyle V_0\, {\rm e}^{-\lambda \phi /M_P}$ &$\displaystyle1 + A_2\,\frac{\phi^2}{2M_P^2}$ \\
\hline
\end{tabular}
\caption{\label{tab:models} Taken from Ref.~\cite{Fischer:2024eic}; list of the quantities defining the three considered STTs; \textit{Chameleon:} different models are distinguished by $n \in \mathbb{Z}^+ \cup 2\mathbb{Z}^-\setminus\{-2\}$. $\Lambda$ and $M_C$ are mass scales that describe the self-interaction of the chameleon and the coupling to matter, respectively. \textit{Symmetron:} The tachyonic mass $\mu$ and the dimensionless self-coupling parameter $\lambda_S$ determine the symmetron potential, while the mass scale $M_S$ sets the matter coupling. \textit{Dilaton:} A constant energy density $V_0$ and a dimensionless self-coupling constant $\lambda$ describe the dilaton potential, and the dimensionless parameter $A_2$ determines the coupling to matter.}
\end{table}
Since we are working with the identification $k(x) = \phi(x)$, Eq.~(\ref{EA2aB}) tells us that, for all considered models, $\Bar{B}(k)=\frac{1}{2}$. This means that, in all three cases, we find SD theories with a canonical kinetic term for $k$. In addition, using Eqs.~(\ref{B3G}) and (\ref{B3Lambda}), we obtain $G(k)$ and $\Lambda(k)$ for each of the models; see Tab.~\ref{tab:SD}. Note that we have employed $A(\phi)-1 \ll 1$ and we have neglected a small term $\sim \lambda_S \phi/M_S$. 
\begin{table}[htbp]
\centering
\renewcommand{\arraystretch}{1.5}
\begin{tabular}{|c||c|c|c|}
\hline\
SD couplings & \textit{Chameleon} & \textit{Symmetron} & \textit{Dilaton} \\
\hline\hline
$G(k)$ & $\displaystyle \frac{1}{M_P^2} e^{\frac{2k}{M_C}}$ & $\displaystyle  \frac{1}{M_P^2} \left(1 +\frac{k^2}{M_S^2} \right)$ & $\displaystyle \frac{1}{M_P^2} \left(1 + A_2\frac{k^2}{M_P^2} \right)$ \\
\hline
$\Lambda(k)$   & $\displaystyle \frac{1}{M_P^2}\frac{\Lambda_C^{4+n}}{k^n} e^{-\frac{2k}{M_C}}$ & $\displaystyle \frac{k^2}{2M_P^2} \left[ -\mu^2 + k^2 \bigg( \frac{\lambda_S}{2} + \frac{\mu^2}{M_S^2}\bigg) \right]$ &
$\displaystyle \frac{V_0}{M_P^2} e^{-\lambda k/M_P} \left( 1- A_2 \frac{k^2}{M_P^2} \right)$
\\
\hline
$\Bar{B}(k)$ & $\displaystyle\frac{1}{2}$ & $\displaystyle\frac{1}{2}$ & $\displaystyle\frac{1}{2}$ \\
\hline
\end{tabular}
\caption{\label{tab:SD} The running couplings of SD theories obtained from three different STTs by using the equivalence relations from Sec.~\ref{sec:ExSTTasSD} }
\end{table}

The form of $k(x)$ is determined by the specific solution of the EOM for $\phi$. For example, the solution for a chameleon around a static, homogeneous sphere with radius $\mathcal{R}$ is given by \cite{Khoury2003}
\begin{eqnarray}   
 \label{csol}
\phi(r) &=& -(\phi_{\text{out}}-\phi_{\text{in}})\frac{\mathcal{R}}{r}e^{-m_{\text{out}}r} + \phi_{\text{out}}~,
\end{eqnarray}
where $\phi_{\text{in}}$ and $\phi_{\text{out}}$ are obtained by evaluating the EOM inside and outside the sphere, respectively, and $m_{\text{out}}$ is the effective mass following from the non-vanishing vacuum expectation value outside the sphere. Consequently, we have 
\begin{eqnarray}   
 \label{ksol}
k(r) &=& -(k_{\text{out}}-k_{\text{in}})\frac{\mathcal{R}}{r}e^{-m_{\text{out}}r} + k_{\text{out}}~.
\end{eqnarray}  
In this particular model, we find from Tab.~\ref{tab:SD} that $G(k)$ is obviously smooth and always larger than $0$. In addition, if we use the scalar field solution in Eq.~(\ref{ksol}), $k$ and $\Lambda(k)$ will also be smooth. Furthermore, the solution in Eq.~(\ref{ksol}) falls off to a constant value $k_{\text{out}}$ exponentially fast. While the scalar potential is bounded from below for even values of $n$, this is not the case for odd positive integers. Moreover, the potential might develop a singularity for these parameter values, depending on the value of $\phi_{\text{out}}$. These properties are then inherited by the cosmological constant $\Lambda(k)$. Since this problem is already present in the STT, it might be necessary to restrict $n$ to even values only for the special case considered here. Overall, for suitable values of $n$, the chameleon example constitutes a clean equivalence pair, demonstrating how a consistent SD theory arises from a realistic STT.

For symmetrons and environment-dependent dilatons, the solutions for $\phi$ around a static, homogeneous sphere are qualitatively similar to Eq.~(\ref{csol}); see Refs.~\cite{Burrage:2016rkv,Brax2022}. Consequently, for this particular setup, and from Tab.~\ref{tab:SD}, we conclude that, in both cases, $G(k)$, $\Lambda(k)$ and $k(x)$ are smooth, $G$ is always larger than $0$, and $k$ falls off exponentially fast, reaching a constant value at infinity. Furthermore, for both models, the scalar potentials and the resulting $\Lambda(k)$ are bounded from below. Therefore, we have shown that also symmetrons and dilatons lead to consistent equivalent SD gravities. Note that the results will likely be very different if another choice for the relation between $\phi(x)$ and $k(x)$ is made.

\subsection{SD gravity models as STTs}

Now, we will also demonstrate how the equivalence relations are applied to various SD models from the literature. In contrast to Sec.~\ref{sec:ExSTTasSD}, where we have worked in the Einstein frame, we will now work in the Jordan frame. Since the models discussed in the literature do not feature a kinetic term for $k$ \cite{Rincon:2019ptp,Alvarez:2022wef,Alvarez:2022mlf,Alvarez:2023ywi,Rincon:2022hpy}, we have $B(k)=0$ in all examples, which means that the resulting STTS will all have scale factors of the form given in Eq.~(\ref{eq:A2SDasSTT}) and only differ by their potentials.

Furthermore, in only one of the examples, the ELFP is imposed, while the others make use of the NEC instead. From our discussion in Sec.~\ref{sec:FieldEqs} we know that SD models, in which the EOM for $k$ is not imposed, cannot result in a consistent STT. Nevertheless, we will apply the equivalence to these models in order to see the results. We will find that the STTs resulting from SD models are significantly less well-behaved than the SD theories obtained from the known STTs in Sec.~\ref{sec:ExSTTasSD}. This is not only due to most SD models not imposing the EOM for $k$, but also because the requirements for a consistent STT are much stricter than those of SD gravity. For example, while the boundedness of the scalar potential is crucial for the compatibility of a STT with observations, it is not unusual for a cosmological constant to diverge in a SD model, e.g., at a cosmological singularity or at the center of a black hole. Since both quantities are connected in the equivalence by Eq.~(\ref{A2aV1}), this will lead to an inconsistency for the STT.


\subsubsection{Einstein-Hilbert cosmology with NEC}

As a first example, we will discuss SD Einstein-Hilbert cosmology in the absence of matter; see, e.g., Refs.~\cite{Canales:2018tbn,Alvarez:2022wef,Alvarez:2022mlf,Alvarez:2020xmk,Hipolito-Ricaldi:2024xlb,Bertini:2024onw}. We assume a homogeneous and isotropic spacetime that is spatially flat. In this case, the only relevant coordinate is the cosmological time $t$ and we can directly work with time-dependent gravitational couplings. The modified Friedmann equations following from Eq.~(\ref{EinsteinSIIIJ}) are
\begin{eqnarray} 
\label{Fr1}
\Lambda(t) &=& -\frac{3\Dot{a}(t)\left(a(t)\Dot{G}(t) - G(t)\Dot{a}(t) \right)}{a(t)^2G(t)}~,
\\
 \label{Fr2}
\Lambda - \frac{\Dot{a}(t)^2}{a(t)^2} &=& \frac{2\Dot{a}(t)\Dot{G}(t)-2G(t)\Ddot{a}(t)}{G(t)a(t)} + \frac{G(t)\Ddot{G}(t) - 2 \Dot{G}(t)^2}{G(t)^2}~,
\end{eqnarray} 
where $a(t)$ is the cosmological scale factor. In order to arrive at a unique solution to these equations, we impose the NEC, which for the effective energy-momentum tensor of SD gravity yields
\begin{eqnarray} 
-2 \left( \frac{\Dot{G}(t)}{G(t)} \right)^2 + \frac{\Ddot{G}(t)}{G(t)} - \frac{\Dot{a}(t)}{a(t)}\frac{\Dot{G}(t)}{G(t)} &=& 0 ~.
\end{eqnarray} 
The solution to the combined system of equations is \cite{Canales:2018tbn}
\begin{eqnarray} 
a(t) &=& a_0 e^{\frac{t}{t_0}}~,~~G(t) = \frac{G_0}{1 + \zeta a(t)}~,~~\Lambda(t) = \Lambda_0 \left( \frac{1 + 2 \zeta a(t)}{1 + \zeta a(t)} \right)~,
\end{eqnarray} 
with $t_0=\pm \sqrt{\frac{3}{\Lambda_0}}$, to which we can apply Eqs.~(\ref{A2aA1}) and (\ref{A2aV1}) to obtain the quantities of the equivalent STT:
\begin{eqnarray} 
A^2(\phi) &=& \frac{M_P^2G_0}{1 + \zeta a_0 e^{\frac{t}{t_0}}}~,~~
V(\phi) = M_P^4 G_0 \Lambda_0  \frac{1 + 2\zeta a_0 e^{\frac{t}{t_0}}}{\left( 1 + \zeta a_0 e^{\frac{t}{t_0}} \right)^2}~.
\end{eqnarray} 
Note that we do not have an expression for $k(x)$ because the system of equations was solved directly for the running couplings. However, this is not a problem since, for $B(k)=0$, $\phi(k)$ is determined via Eq.~(\ref{A2a3})
\begin{eqnarray} 
\label{eq:ExplResphi}
\phi(t) &=& \pm \sqrt{\frac{3}{2}}M_P \ln G(t) = \pm \sqrt{\frac{3}{2}}M_P \ln \left( \frac{G_0}{1 + \zeta a(t)} \right)~.
\end{eqnarray} 
By inverting Eq.~(\ref{eq:ExplResphi}) for $\phi$, we can express $G(t)$ as well as $A^2(\phi)$ and $V(\phi)$ in terms of $\phi$. It is then straightforward to verify that $\phi$ does not satisfy its EOM (\ref{EOMphi}), as expected from the discussion in Sec.~\ref{sec:FieldEqs}. 
We see that all quantities of the STT fulfill the conditions discussed in Sec.~\ref{sssec:SDtoSTT}. In addition, we have $V(\phi) > 0$, i.e., the scalar potential is bounded from below. However, $\phi$ does not satisfy any fall-off conditions, but diverges for $t \rightarrow \infty$, which likely is a consequence of the field not satisfying its EOM.


\subsubsection{Einstein-Hilbert cosmology with ELFP}

As a second example, we investigate the SD bouncing cosmology model of Ref.~\cite{Alvarez:2023ywi}. Again, the only relevant coordinate is the cosmological time $t$ and we will directly work with time-dependent gravitational couplings. In bouncing cosmologies, the NEC is automatically violated, which is why the ELFP is employed to uniquely fix the theory. It is important to note that the condition imposed in Ref.~\cite{Alvarez:2023ywi} is actually stronger than the ELFP given by Eq.~(\ref{ELFP}). More precisely, it is demanded that Eq.~(\ref{ELFP}) is satisfied for any $k(t)$, which is only possible for a constant Lagrangian. In this situation, the EOM for $k$ trivializes. Since the Lagrangian of the equivalent STT will also be constant, the EOM for $\phi$ is trivial as well and no inconsistencies of the equivalence arise at the level of the field equations. Again using Eqs.~(\ref{Fr1}) and (\ref{Fr2}), we complement them by the stronger version of the ELFP:
\begin{eqnarray}
\frac{6\left( \Dot{a}(t)^2 + a(t) \Ddot{a}(t) \right)}{a(t)^2 G(t)} - 2\frac{\Lambda(t)}{G(t)} &=& c_L~,
\end{eqnarray}
where $c_L$ is some constant. The resulting system of equations can be solved analytically:
\begin{eqnarray}
&&a(t) = c_2 \left( \cosh \left[ \sqrt{\frac{c_R}{3}}(c_1 + t) \right] \right)^{\frac{1}{2}}~,
\\
&&G(t) 
= \Bigg\{ \frac{\mathcal{F}(t_i; c_R,c_1)}{\mathcal{F}(t; c_R,c_1)} \Bigg[ \frac{1}{G_i} - \frac{c_L}{|c_R|} + \frac{c_L}{|c_R|} \frac{\mathcal{F}(t; c_R,c_1)}{\mathcal{F}(t_i; c_R,c_1)} 
\nonumber
\\
&&
~~~~~~~~~
+ \frac{ic_L}{|c_R|\mathcal{F}(t_i; c_R,c_1)} \left( E^{(F)}(i \sqrt{\frac{c_R}{12}}(t+c_1),2) - E^{(F)}(i \sqrt{\frac{c_R}{12}}(t_i + c_1),2) \right)  \Bigg]  \Bigg\}^{-1}~,
\\
&&\Lambda(t) = \frac{1}{2} \left( c_R - c_L G(t) \right)~, 
\end{eqnarray}
where
\begin{eqnarray}
\mathcal{F} &=& \frac{\sqrt{\cosh \left( \sqrt{\frac{c_R}{3}}(t + c_1) \right)}}{\sinh \left( \sqrt{\frac{c_R}{3}}(t + c_1) \right)}
~,~~
E^{(F)}(x,m) = \int_{0}^{x}d\theta \frac{1}{\sqrt{1-m \sin^2(\theta)}}~,
\end{eqnarray}
and $c_1 = -t_b$, $c_R = \Tilde{R}_0 = 4 \Lambda_{0}$ and $c_2 = \sqrt{2}a_0 e^{\sqrt{\frac{\Lambda_0}{3}}t_b}$ with $t_b$ being the bouncing time, $\Tilde{R}_0$ being the constant Ricci scalar, which is related to the observed value of the cosmological constant $\Lambda_{0}$, and $a_0$ being related to the scale factor. $G_i$ is another integration constant. For more details we refer to Ref.~\cite{Alvarez:2023ywi}. Due to the complicated form of $G(t)$, we refrain from writing it out explicitly from now on. For this model, Eqs.~(\ref{A2aA1}), (\ref{A2aV1}) and (\ref{A2a3}) give
\begin{eqnarray}
A^2(\phi) = M_P^2G(t)~,~~
V(\phi) =  2M_P^4\Lambda_0 G(t) - \frac{M_P^4}{2}c_L G(t)^2~,~~
\phi(t) = \pm \sqrt{\frac{3}{2}}M_P \ln G(t)~.
\end{eqnarray}
Again, inverting the last equation, allows us to express $A^2$ and $V$ in terms of $\phi$. 

We will now apply the discussion in Ref.~\cite{Alvarez:2023ywi} of the functional behavior of $G(t)$ to the equivalent STT. While the specific properties of $G$ crucially depend on the parameter choice, there are some general features that can be interpreted in the present context. Most importantly, at the bouncing time $t=t_b$, $G(t)$ actually diverges towards positive infinity from the right and to negative infinity from the left. The latter violates our requirement that $G(t)>0$, our assumption about smoothness as well as the condition $B(k) > - \frac{3}{4}\frac{1}{G(k)^3}\left( \frac{\partial G(k)}{\partial k} \right)^2$. Due to the dependence of $\phi$, $A^2$ and $V$ on $G$, this affects the whole STT. The scalar potential cannot be bounded from below, and, since $G(t)$ falls off to zero for very early and late times, $\phi$ diverges at infinity. Even worse, $\phi$ is given as the logarithm of $G$, which is not a real number for negative arguments. Overall, while the ELFP indeed allows us to arrive at a STT in which $\phi$ satisfies its EOM, the criteria of the equivalence are violated and it is not possible to define a consistent STT corresponding to this particular SD model.


\subsubsection{Polytropic black hole with NEC}

In the last example, we will deal with the polytropic black hole solution obtained in Ref.~\cite{Contreras:2018dhs}. Due to the fact that this is a static, spherically symmetric solution, the only relevant variable is the radius $r$. Applying the NEC (\ref{NEC}) to the effective energy-momentum tensor of SD gravity gives
\begin{eqnarray}
\label{polyGeq}
\left( \frac{\partial G(r)}{\partial r} \right)^2 &=& \frac{1}{2}G(r) \frac{\partial^2 G(r)}{\partial r^2}~,
\end{eqnarray}
which is solved by
\begin{eqnarray}
\label{eq:Gpolytrop}
G(r) &=& \frac{G_0}{1+\epsilon r}~,
\end{eqnarray}
where $\epsilon$ is an integration constant chosen such that $G(r)$ coincides with the classical value $G_0$ for $\epsilon \rightarrow 0$. Eq.~(\ref{eq:Gpolytrop}) is then used to solve the modified Einstein equation with a contribution to the energy-momentum tensor coming from matter satisfying a polytropic equation of state. We are only interested in the solution for the cosmological constant
\begin{eqnarray}
\Lambda(r) &=& f_0(r)\Lambda_{0} - f_1(r) + f_2(r) \ln \left( 2 M_0 G_0 \frac{\epsilon r + 1}{r} \right)~,
\end{eqnarray}
with
\begin{eqnarray}
f_0(r) &=& \frac{(2\epsilon r + 1)}{\epsilon r + 1}
~,~~
f_1(r) = \frac{3M_0 G_0}{8\pi} \frac{(12 \epsilon r (\epsilon r + 1) + 1)\epsilon^2}{r(\epsilon r + 1)^2}
~,~~
f_2(r) = \frac{18 M_0 G_0}{8 \pi} \frac{\epsilon^3(2 \epsilon r +1)}{\epsilon r +1}~.~~~~~~
\end{eqnarray}
Consequently, Eqs.~(\ref{A2aA1}), (\ref{A2aV1}) and (\ref{A2a3}) lead us to
\begin{eqnarray}
A^2(\phi) &=& \frac{M_P^2 G_0}{1 + \epsilon r}
~,~~
V(\phi) = \frac{M_P^4 G_0 }{1 + \epsilon r} \left[ \Lambda_0 f_0(r) - f_1(r) + f_2(r) \ln \left( 2 M_0 G_0 \frac{\epsilon r + 1}{r} \right) \right]
~,
\nonumber
\\
\phi(r) &=& \pm \sqrt{\frac{3}{2}}M_P \ln G(r)~.
\end{eqnarray}
Solving the last equation for $r$, we can then express $A^2$ and $V$ in terms of $\phi$.

Apart from the usual singularity at $r=0$, all quantities of the resulting STT are smooth. Further, all the other conditions discussed in Sec.~\ref{sssec:SDtoSTT} are satisfied. Regarding the sanity of the theory, since the EOM for $k$ was not imposed, $\phi$ cannot satisfy its own EOM. The scalar field also does not satisfy any fall-off conditions, but instead diverges for $r \rightarrow \infty$. In addition, $V$ is unbounded from below as it diverges to negative infinity for $r \rightarrow 0$. This behavior of the scalar potential does not come as a surprise since, in the context of the equivalence, the potential corresponds to the cosmological constant which approaches negative infinity at the singularity of the black hole, signaling a breakdown of the theory.


\section{Conclusions}
\label{sec:Conclusion}

In order to tackle striking issues in modern cosmology like dark matter or dark energy, modified theories of gravity are frequently investigated. While STTs and $f(R)$-gravity likely are the most well-known approaches, there also exist other ideas like SD gravity, which is a low-energy effective theory inspired by asymptotic safety. Studying how such modified gravity theories are related to each other may tell us more about the nature of gravity itself. Therefore, in this article, we have successfully established an equivalence between SD gravity and STTs with a single scalar field, a canonical kinetic term and conformal coupling to the metric tensor. We found that, on the level of the action, any SD theory with smooth running couplings, $G(k) > 0$, as well as $B(k) > - \frac{3}{4}\frac{1}{G(k)^3}\left( \frac{\partial G(k)}{\partial k} \right)^2$ can be uniquely rewritten as a STT. This particularly includes all well-behaved scale-dependent models discussed in the literature, see Ref.~\cite{Rincon:2019ptp} and references therein, proving that this class of theories can be fully embedded into STTs. Regarding the other direction, there are multiple ways of mapping a STT to an equivalent SD theory, differing in the definitions of the scale-setting relation and the prefactor of the kinetic term for $k$. In this case, the equivalence is applicable to practically any STT whose action is of the form given in Eq.~(\ref{SIJ}). On the level of the field equations, we found that there is an intimate connection between the EOMs for $\phi$ and $k$. In a SD theory resulting from a STT, any $k(\phi)$ will automatically satisfy its EOM. On the other hand, if we start with a SD model in which $k$ does not satisfy its EOM, the scalar field of the equivalent STT will also not satisfy its EOM and must, therefore, be treated as a non-dynamical entity, which is, however, not in the spirit of STTs. This connection between the EOMs lends further support to the idea of promoting $k$ to a dynamical entity with its own kinetic term and EOM. It is important to note that the EOM for $k$ also results from demanding that, in accordance with the Bianchi identities, the divergence of the effective energy-momentum tensor vanishes. In Ref.~\cite{Reuter:2003ca}, it was shown that by suitably choosing the running couplings and complementing the action with a term containing derivatives of the Newton coupling, a vanishing divergence of the effective energy-momentum tensor can be achieved without imposing the EOM for $k$. However, as we have demonstrated in this article, this additional term in the action is nothing but a kinetic term for $k$ with a specific choice of $B(k)$. By simply treating $k$ as a dynamical field and imposing its EOM, the same result can be achieved for arbitrary running couplings and an arbitrary kinetic term, including $B(k)=0$.

By mapping $f(R)$-gravity onto itself in Sec.~\ref{sec:Consistency}, we have provided further evidence for the consistency of the equivalence. In this way, we have also illustrated that the new equivalence fits perfectly into the framework of existing ones involving $f(R)$-theory, thus completing the equivalence triangle in Fig.~\ref{fig:EquiTri1}. In Sec.~\ref{sec:Examples}, we have also shown that the equivalence relations can easily be applied to concrete models of SD gravity and STTs. The considered screened scalar field models led to well-behaved SD theories, which are, however, not of the type usually discussed in the literature since they contain a kinetic term for $k$. This means that we can currently not apply the existing experimental parameter constraints obtained for screened scalar models to SD gravity. In contrast, the STTs resulting from the considered SD gravity examples are far less well-behaved, which is mostly due to the fact that the original theories are not well-behaved themselves.

\begin{figure}[htbp]
\begin{center}
\includegraphics[scale=0.8]{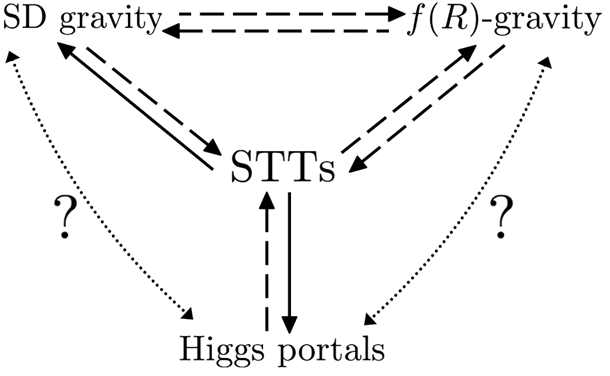}
\caption{The equivalence triangle from Fig.~\ref{fig:EquiTri1} amended by equivalences to Higgs portal models \cite{Schabinger:2005ei,Patt:2006fw}; Ref.~\cite{Burrage:2018dvt} suggests that every STT of the type considered in the present article can be expressed as a Higgs portal theory, while not every Higgs portal theory is expected to be equivalent to a STT. From this figure, we can conclude that every SD theory with smooth couplings $G(k) >0$ and $B(k)$ fulfilling Eq.~(\ref{eq:third}) should be equivalent to Higgs portal models. However, the direct relations between Higgs portals and SD gravity or $f(R)$-gravity remain to be studied.}
\label{fig:EquiTri2}
\end{center}
\end{figure}

There are some interesting directions that can now be explored. In this work, we have focused mainly on analyzing the equivalence itself, while we have only touched on its applications. Now that the equivalence triangle is completed, any method or result from one theory may be applied to the other two, potentially leading to various new insights about modified gravity in general. Especially, we hope that the relation between the dynamics of the scale-setting relation and the scalar field depicted here motivates the study of SD models, in which $k$ is treated as a dynamical entity. A starting point might be the generalization of the ELFP to the full EOM for $k$ and the subsequent application of this condition to different scenarios, such as cosmology and black holes. Another promising avenue might be to incorporate the equivalence between STTs and certain Higgs portal theories described in Ref.~\cite{Burrage:2018dvt} into the equivalence picture; see Fig.~\ref{fig:EquiTri2}. Since there exist equivalences of STTs with all the aforementioned theories, it can be anticipated that $f(R)$-gravity and SD gravity should also be equivalent to Higgs portal models. 
Note that there also exist other investigations of scalar-tensor theories in different conformal frames and parameterizations \cite{Jarv:2014hma}, which have addressed the GR limit \cite{Jarv:2015kga}, inflation \cite{Jarv:2016sow} and a generalization to the Palatini formalism \cite{Jarv:2020qqm,Jarv:2024krk}. It would be interesting to discuss how these studies can be connected to the results obtained in the present article.
Last, there is a crucial aspect of the equivalence that needs further investigation. Even though the action in Eq.~(\ref{SIIIJ}) is only the lowest order part of the full SD action, it is well-known that this action is a full quantum result and our ignorance of the concrete UV-completion of gravity is incorporated into the running of the gravitational couplings. In sharp contrast, STTs are classical field theories, ignorant of the quantum nature of gravity \cite{Laporte:2021kyp}. This now raises the question of how an effective quantum gravity theory can be equivalent to a classical field theory, at least at the lowest level of perturbation. Answering this question could lead to new insights into the quantum nature of gravity and STTs.

\begin{acknowledgments}

The authors are grateful to C.~Burrage, H.~Fischer, P.~Millington, R.~I.~P.~Sedmik and M.~Shams~Nejati for useful discussions. This research was funded in whole or in part by the Austrian Science Fund (FWF) [10.55776/PAT7599423] and [10.55776/PAT8564023], and is based upon work from COST Action COSMIC WISPers CA21106, supported by COST (European Cooperation in Science and Technology). For open access purposes, the author has applied a CC BY public copyright license to any author accepted manuscript version arising from this submission.
\end{acknowledgments}


\appendix


\section{Modified Einstein equations in the Jordan frame}
\label{app:ModEinsEqs}

Here, we will derive the modified Einstein equations for SD gravity that are used in Sec.~\ref{sec:FieldEqs}. Varying the action in Eq.~(\ref{SIIIJ}) with respect to the metric yields
\begin{eqnarray}
\label{varSIIIJ}
\delta \Tilde{S}_\text{SD} &=& \int d^4x \left[ \delta \sqrt{-\Tilde{g}}\left( \frac{\Tilde{R}-2\Lambda(k)}{2G(k)} -B(k) (\Tilde{\partial}k)^2 \right) + \sqrt{-\Tilde{g}} \delta \left( \frac{\Tilde{R}}{2G(k)} \right) -\sqrt{-\Tilde{g}}B(k)\delta (\Tilde{\partial}k)^2 \right]~.~~~~~~
\end{eqnarray}
Note that $\delta \Tilde{g}_{\mu \nu} = -\Tilde{g}_{\alpha \mu}\Tilde{g}_{\beta \nu}\delta \Tilde{g}^{\alpha \beta}$ and $\delta \sqrt{-\Tilde{g}} = -\frac{1}{2}\sqrt{-\Tilde{g}}\Tilde{g}_{\mu \nu}\delta \Tilde{g}^{\mu \nu}$. The variation of the Ricci tensor is given by the Palatini identity
\begin{eqnarray}
\delta \Tilde{R}_{\mu \nu} &=& \Tilde{\nabla}_{\alpha}\delta \Tilde{\Gamma}^{\alpha}_{\mu \nu}-\Tilde{\nabla}_{\nu}\delta \Tilde{\Gamma}^{\alpha}_{\alpha \mu}~,
\end{eqnarray}
from which we obtain
\begin{eqnarray}
\label{eq:varRicci1}
\delta \Tilde{R} &=& \Tilde{R}_{\mu \nu}\delta \Tilde{g}^{\mu \nu} + \Tilde{\nabla}_{\alpha}\left( \Tilde{g}^{\mu \nu}\delta\Tilde{\Gamma}^{\alpha}_{\mu \nu} - \Tilde{g}^{\mu \alpha}\delta\Tilde{\Gamma}^{\nu}_{\nu \mu} \right)~.
\end{eqnarray}
Note that $\delta \Tilde{R}$ is multiplied by $\frac{\sqrt{-\Tilde{g}}}{2G(k(x))}$ in Eq.~(\ref{varSIIIJ}). In GR, $G$ is constant and the factor $\sqrt{-\Tilde{g}}$ is not affected by the covariant derivative of the second term on the right-hand side of Eq.~(\ref{eq:varRicci1}), which is then a total derivative and can usually be discarded since it does not change the EOM. However, in SD gravity, the situation is different: $G$ is no longer constant and the second term is not a total derivative. Writing out the Christoffel symbols gives
\begin{eqnarray}
\label{eq:varRicci2}
\delta \Tilde{R} &=& (-\Tilde{R}^{\mu \nu} - \Tilde{\nabla}^{\mu}\Tilde{\partial}^{\nu}+\Tilde{g}^{\mu \nu} \Tilde{\nabla}^{\alpha}\Tilde{\partial}_{\alpha})\delta \Tilde{g}_{\mu \nu}~.
\end{eqnarray}
We see that the second term on the right-hand side of Eq.~(\ref{eq:varRicci2}) is actually proportional to the variation of the derivative of the metric. For this reason, in the Euler-Lagrange equations we would obtain a non-zero term $\frac{\partial \Tilde{L}_\text{SD}}{\partial (\partial_{\alpha} g_{\mu \nu})}$ that changes the Einstein equations. A strategy to deal with the second term on the right-hand side of Eq.~(\ref{eq:varRicci2}) is to use partial integration to move all derivatives away from the variation of the metric. These derivatives will then, in Eq.~(\ref{varSIIIJ}), instead act on $\frac{\sqrt{-\Tilde{g}}}{2G(k(x))}$. More precisely, we have 
\begin{eqnarray}
\sqrt{-\Tilde{g}}\frac{1}{2G(k)}\Tilde{\nabla}^{\mu}\Tilde{\partial}^{\nu}\delta \Tilde{g}_{\mu \nu}
&=& \sqrt{-\Tilde{g}} \frac{1}{2G(k)^2}\Tilde{\nabla}^{\mu}G(k)\Tilde{\partial}^{\nu} \delta \Tilde{g}_{\mu \nu} 
\nonumber
\\
&=& - \sqrt{-\Tilde{g}}\Tilde{\nabla}^{\nu}\left(\frac{1}{2G(k)^2} \Tilde{\nabla}^{\mu}G(k) \right) 
\nonumber
\\
&=& \sqrt{-\Tilde{g}}\left( \frac{1}{G(k)^3} \Tilde{\nabla}^{\nu}G(k) \Tilde{\nabla}^{\mu}G(k) - \frac{1}{2G(k)^2} \Tilde{\nabla}^{\nu} \Tilde{\nabla}^{\mu} G(k)  \right)\delta \Tilde{g}_{\mu \nu}~, 
\end{eqnarray}
where in the second-to-last line we have used that for any vector $Z^{\mu}$:
\begin{eqnarray} 
\label{vectordensity}
\Tilde{\partial}_{\mu}(\sqrt{-\Tilde{g}}Z^{\mu}) &=& \Tilde{\nabla}_{\mu}(\sqrt{-\Tilde{g}}Z^{\mu})~.
\end{eqnarray}
Furthermore, we have
\begin{eqnarray} 
-\sqrt{-\Tilde{g}}\Tilde{g}^{\mu \nu} \frac{1}{2G(k)}\Tilde{\nabla}^{\alpha}\Tilde{\partial}_{\alpha}\delta \Tilde{g}_{\mu \nu} 
&=& \sqrt{-\Tilde{g}}\Tilde{g}^{\mu \nu}\left( -\frac{1}{G(k)^3}(\Tilde{\nabla}G(k))^2+\frac{1}{2G(k)^2}\Tilde{\Box}G(k) \right)\delta \Tilde{g}_{\mu \nu}~.
\end{eqnarray}
Therefore, the variation of the full action is
\begin{eqnarray} 
\label{varSIIIJfull}
\delta \Tilde{S}_\text{SD} 
&=& 
\int d^4x \sqrt{-\Tilde{g}} \frac{1}{2G(k)} \left[ \Tilde{R}_{\mu \nu}-\frac{1}{2}\Tilde{R}\Tilde{g}_{\mu \nu} + \Lambda(k)\Tilde{g}_{\mu \nu} + \Delta_{G}t_{\mu \nu} + \Delta_{B}t_{\mu \nu} \right]\delta \Tilde{g}^{\mu \nu}~,
\end{eqnarray}
where $\Delta_{G}t_{\mu \nu}$ and $\Delta_{B}t_{\mu \nu}$ are defined in Eqs.~(\ref{DtG}) and (\ref{DtB}). From this, we then find the modified Einstein equations \ref{EinsteinSIIIJ}. The derivation of the modified Einstein equations for STTs, see Eq.~(\ref{EinsteinSIJ}), is completely analogous to the above calculation. 


\section{Divergence of the modified energy-momentum tensor}
\label{app:Divergence}

Here, we give explicit expressions for the individual terms in Eq.~(\ref{newdiv}):
\begin{eqnarray}    
-\Tilde{G}_{\mu \nu} \Tilde{\nabla}^{\mu}\left( \frac{1}{2G(k)} \right) &=& \Tilde{R}_{\mu \nu} \frac{1}{2G(k)^2} \left( \frac{\partial G(k)}{\partial k} \right) \Tilde{\nabla}^{\mu}k - \Tilde{R} \frac{1}{4G(k)^2} \left( \frac{\partial G(k)}{\partial k} \right) \Tilde{\nabla}_{\nu}k~,
\\
-\Tilde{\nabla}^{\mu}\left( \frac{1}{2G(k)} \Lambda(k)\Tilde{g}_{\mu \nu} \right) &=& \Lambda(k)\frac{1}{2G(k)^2}\left( \frac{\partial G(k)}{\partial k} \right)\Tilde{\nabla}_{\nu}k - \frac{\partial \Lambda(k)}{\partial k}\frac{1}{2G(k)}\Tilde{\nabla}_{\nu}k~,
\\
-\Tilde{\nabla}^{\mu} \left(\frac{1}{2G(k)^2}\Delta_{B}t_{\mu \nu}  \right) &=& \frac{1}{2} \frac{\partial B(k)}{\partial k} (\Tilde{\partial}k)^2\Tilde{\nabla}_{\nu}k + B(k)\Tilde{\Box}k \Tilde{\nabla}_{\nu}k~,
\\
\label{divterm4}
-\Tilde{\nabla}^{\mu} \left(\frac{1}{2G(k)^2}\Delta_{G}t_{\mu \nu}  \right) &=& \frac{1}{2G(k)^2}\left( \frac{\partial G(k)}{\partial k} \right) \left[\Tilde{\nabla}_{\nu}\Tilde{\Box}k-\Tilde{\nabla}_{\mu}\Tilde{\nabla}_{\nu}\Tilde{\nabla}^{\mu}k  \right]~,
\end{eqnarray}
where we have used that $[\Tilde{\nabla}_{\mu},\Tilde{\nabla}_{\nu}]f(x)=0$ for any scalar function $f(x)$ in the absence of torsion. Using the definition of the Riemann tensor, Eq.~(\ref{divterm4}) can be written as
\begin{eqnarray} 
-\Tilde{\nabla}^{\mu} \left(\frac{1}{2G(k)^2}\Delta_{G}t_{\mu \nu}  \right) 
&=& \frac{1}{2G(k)^2}\left( \frac{\partial G(k)}{\partial k} \right) [\Tilde{\nabla}_{\nu},\Tilde{\nabla}_{\mu}]\Tilde{\nabla}^{\mu}k 
\nonumber
\\
&=& \frac{1}{2G(k)^2}\left( \frac{\partial G(k)}{\partial k} \right)\Tilde{R}^{\mu}_{\ \alpha \nu \mu}\Tilde{\nabla}^{\alpha}k 
\nonumber
\\
&=& -\frac{1}{2G(k)^2}\left( \frac{\partial G(k)}{\partial k} \right) \Tilde{R}_{\mu \nu}\Tilde{\nabla}^{\mu}k~.
\end{eqnarray}


\bibliography{refs}

\end{document}